\documentstyle[12pt,epsf]{article}

\newcommand{\dlt}{\delta}
\newcommand{\Dlt}{\Delta}

\newcommand{\gm}{\gamma}
\newcommand{\Gm}{\Gamma}
\newcommand{\tht}{\theta}
\newcommand{\btht}{\bar{\tht}}
\newcommand{\lmd}{\lambda}
\newcommand{\Lmd}{\Lambda}
\newcommand{\sgm}{\sigma}
\newcommand{\vph}{\varphi}

\newcommand{\be}{\begin{equation}}
\newcommand{\ee}{\end{equation}}
\newcommand{\bea}{\begin{eqnarray}}
\newcommand{\eea}{\end{eqnarray}}
\newcommand{\eql}{\!\!\!&=\!\!\!&}

\newcommand{\defa}{\!\!\!&\equiv\!\!\!&}

\newcommand{\mtrx}[4]{\brkt{\begin{array}{cc}#1&#2\\#3&#4\end{array}}}
\newcommand{\vct}[2]{\brkt{\begin{array}{c}#1\\#2\end{array}}}

\newcommand{\tl}[1]{\tilde{#1}}
\newcommand{\bdm}[1]{{\mbox{\boldmath $#1$}}}

\newcommand{\diag}{{\rm diag}}
\newcommand{\der}{\partial}
\newcommand{\dr}{\!\!{\rm d}}

\newcommand{\Acl}{A_{\rm cl}}
\newcommand{\vpcl}{\varphi_{\rm cl}}

\newcommand{\brkt}[1]{\left( #1 \right)}
\newcommand{\brc}[1]{\left\{ #1 \right\}}

\renewcommand{\Re}{{\rm Re}}
\renewcommand{\Im}{{\rm Im}}

\newcommand{\cA}{{\cal A}}
\newcommand{\cD}{{\cal D}}
\newcommand{\cF}{{\cal F}}

\newcommand{\cL}{{\cal L}}

\newcommand{\cN}{{\cal N}}
\newcommand{\cO}{{\cal O}}

\newcommand{\NP}[1]{{\it Nucl.~Phys.}~{\bf #1}}
\newcommand{\PL}[1]{{\it Phys.~Lett.}~{\bf #1}}

\newcommand{\PR}[1]{{\it Phys.~Rev.}~{\bf #1}}
\newcommand{\PRL}[1]{{\it Phys.~Rev.~Lett.}~{\bf #1}}
\newcommand{\PTP}[1]{{\it Prog.~Theor.~Phys.}~{\bf #1}}

\begin{document}

\begin{titlepage}
\null
\begin{flushright}
 {\tt hep-th/0207159}\\
TU-663
\\
July 2002
\end{flushright}

\vskip 2cm
\begin{center}
{\LARGE \bf  Superfield Description of Effective Theories \\
on BPS Domain Walls}

\lineskip .75em
\vskip 2.5cm

\normalsize

{\large \bf Yutaka Sakamura}
{\def\thefootnote{\fnsymbol{footnote}}
\footnote[5]{\it  e-mail address:
sakamura@tuhep.phys.tohoku.ac.jp}}

\vskip 1.5em

{\it Department of Physics, Tohoku University\\ 
Sendai 980-8578, Japan}

\vspace{18mm}

{\bf Abstract}\\[5mm]
{\parbox{13cm}{\hspace{5mm} \small
We derive the low-energy effective theory on the BPS domain wall in 4D $\cN=1$ 
global SUSY theories in terms of the 3D superfields. 
Our derivation makes the preserved SUSY by the wall manifest 
and the procedure for integrating out the massive modes easier. 
Our procedure clarifies how the 3D superfields are embedded into 
the 4D chiral and vector superfields. 
We also point out a shortcoming of the conventional procedure 
for deriving the effective theory on the wall. 
}}

\end{center}

\end{titlepage}

\clearpage

\section{Introduction}
Domain walls are inherent to the field theories with spontaneous 
breaking of a discrete symmetry. 
There are extensive researches on domain walls in many areas of physics,  
such as condensed matter physics, thermal evolution of the universe, 
chiral theories on the lattice, and so on.

In supersymmetric (SUSY) theories, the domain wall sector 
breaks not only the translational invariance but also 
supersymmetry of the original theory. 
Some domain walls, however, preserve part of the original supersymmetry. 
They saturate the Bogomol'nyi-Prasad-Sommerfield (BPS) bound~\cite{BPS}, 
and are called {\it BPS domain walls}\footnote{
The first example of the BPS domain wall was given in Ref.\cite{Cvetic}. 
}. 
BPS saturated states like the BPS domain walls play a crucial role 
in the quantum field theories because their mass spectrum receives 
no quantum corrections. 
For instance, they can be used as powerful tools 
for the investigation of the vacuum structures at strong coupling regime. 
In addition, from the phenomenological point of view, 
the BPS domain walls are important in the brane-world 
scenario~\cite{Rubakov,Arkani,RS} 
because they are stable and can provide a natural realization 
of the partial SUSY breaking of the minimal five-dimensional (5D) SUSY 
(eight supercharges) to $\cN=1$ SUSY (four supercharges), 
which is relevant to the phenomenology. 

For the above reasons, the BPS domain wall 
is an intriguing subject for the study of the field theory. 
In particular, BPS domain walls in four-dimensional (4D) $\cN=1$ theories 
are thoroughly researched in a number of papers 
because such theories are tractable 
and have various types of BPS domain walls 
with interesting features~\cite{Dvali,Chibisov,Gabadadze,HLS,Naganuma}. 
For example, the authors of Ref.\cite{Dvali,Chibisov} discussed 
a BPS domain wall in a simple Wess-Zumino model in detail  
and derived the low-energy effective theory (LET). 
Surveying LET in the BPS wall background 
is a useful approach to investigate the quantum fluctuation of 
the BPS domain wall. 
Since the zero-modes are localized on the wall, 
LET becomes a theory {\it on the wall}. 
In other words, domain wall backgrounds gives rise to 
some kind of the dimensional reduction. 
The authors of Ref.\cite{Dvali} referred to it 
as ``dynamical compactification'',
in contrast to the naive Kaluza-Klein compactification\cite{Klein}. 
Thus, studying LET on the wall background 
is also useful for the purpose of the model-building 
in the brane-world scenario. 
Of course, since our world is four-dimensional, we should discuss 
a domain wall in 5D theories for the realistic model-building.  
However, 5D SUSY theory is quite restrictive due to $\cN=2$ SUSY, 
and difficult to handle. 
Hence, it is convenient and instructive to study the BPS walls 
in 4D $\cN=1$ theories as a toy model. 

In this paper, we will discuss LET on 
the BPS domain wall in 4D $\cN=1$ theories. 
Since BPS walls preserve a half of the original SUSY, 
such LETs become 3D $\cN=1$ theories. 
This means that LET on the BPS wall can be described 
in terms of the 3D superfields. 

However, a conventional procedure for deriving LET on the BPS wall, 
which was discussed intensively in Ref.\cite{Dvali,Chibisov},
does not respect the preserved SUSY by the wall, 
and the resulting LET is described by the component fields 
of the 3D supermultiplets without the auxiliary fields. 
Although such a conventional procedure 
can be used in the computation of the mass spectrum in LET, 
a difficulty arises when we try to calculate interaction terms in LET. 
Namely, such a procedure is inconvenient for integrating out 
the massive modes in order to derive LET. 
We will explain this problem in the next section. 

The above problem comes from the fact that the conventional procedure 
does not keep SUSY preserved by the wall to be manifest. 
So we need an alternative procedure for deriving LET  
on the BPS wall where the preserved SUSY is manifest. 
This is the purpose of the present paper. 

Recall that the actions of 4D $\cN=1$ theories are expressed 
by the form of the integration over the 4D $\cN=1$ superspace 
$(x^m,x_2,\tht_1,\tht_2)$, where $x^m$ ($m=0,1,3$) denote 
the 3D coordinates on the wall, $x_2$ is the coordinate 
of the extra dimension, $\tht_1$ and $\tht_2$ are the fermionic coordinates 
for the broken and unbroken SUSY, respectively. 
Since LET derived by the conventional procedure has the form of 
the integration over the 3D space-time, 
we can say that the conventional procedure corresponds to the execution 
of the explicit integrations in terms of $x_2$, $\tht_1$ and $\tht_2$. 
On the other hand, we want to leave the 3D superspace~$(x^m,\tht_2)$ 
not to be integrated 
in order to make the unbroken SUSY manifest. 
Thus, our desirable procedure corresponds to the execution of 
the explicit integration in terms of {\it only coordinates 
for the broken symmetries}, that is, $x_2$ and $\tht_1$. 
However, due to the complexity of the dependence of the integrands on 
the fermionic coordinates for the broken and unbroken SUSYs, 
such integrations are not easy to be carried out. 
In this paper, we will perform such integrations systematically 
and derive LET on the BPS wall that is described by the 3D superfields. 

The paper is organized as follows. 
In the next section, we will review the conventional procedure 
for the derivation of LET on a BPS wall, 
and point out its shortcoming. 
In Section~\ref{embedding}, we will find how 3D scalar superfields 
are embedded into a 4D chiral superfield.  
Using the result of Section~\ref{embedding}, 
LET on the BPS wall can be derived in the case 
of the generalized Wess-Zumino model. 
The detailed derivation is explained in Section~\ref{3D_LET_deriv}. 
We will also discuss the model with gauge supermultiplets 
in Section~\ref{vector_SF}. 
Section~\ref{summary} is devoted to the summary and the discussion. 
Notations and some useful formulae are listed in the appendices.

\section{Conventional derivation of LET on a BPS wall} \label{motivation}
To illustrate a shortcoming of the conventional procedure 
for deriving LETs on BPS walls, 
let us consider a simple Wess-Zumino model. 
The Lagrangian of the model is\footnote{
Basically, we will follow the notations of Ref.\cite{WB} 
throughout the paper. 
} 
\be
 \cL=\int\dr^2\tht{\rm d}^2\btht\; \bar{\Phi}\Phi
 +\int\dr^2\tht\; W(\Phi)+\int\dr^2\btht\;\bar{W}(\bar{\Phi}), 
\ee
where 
\be
 W(\Phi)=\Lmd^2 \Phi-\frac{g}{3}\Phi^3, \;\;\; (\Lmd,g>0)
\ee
and
\be
 \Phi(y,\tht)=A(y)+\sqrt{2}\tht\Psi(y)+\tht^2F(y), 
\ee
where $y^\mu=x^\mu+i\tht\sgm^\mu\btht$. 

After eliminating the auxiliary field $F$ by the equation of motion, 
the Lagrangian becomes 
\be
 \cL=-\der^\mu\bar{A}\der_\mu A-i\bar{\Psi}\bar{\sgm}^\mu\der_\mu \Psi
 +gA\Psi^2+g\bar{A}\bar{\Psi}^2-|\Lmd^2-g A^2|^2. 
 \label{on_shell_L}
\ee

This theory has the following classical field configuration 
as a solution of the equation of motion. 
\bea
 \Acl(x_2) \eql \frac{\Lmd}{\sqrt{g}}\tanh(\sqrt{g}\Lmd x_2), \nonumber\\
 \Psi_{\rm cl}(x_2) \eql 0.
\eea
This is the BPS domain wall configuration in this theory\footnote{
In this paper, we will choose the $x_2$-direction to be perpendicular 
to the wall.  
}. 

Next, we will consider the fluctuation fields around the above 
classical configuration~$\Acl$ and $\Psi_{\rm cl}$. 
\bea
 A(x) \eql \Acl(x_2)+\frac{1}{\sqrt{2}}(a(x)+ib(x)), \nonumber\\
 \Psi(x) \eql \frac{1}{\sqrt{2}}(\psi_1(x)+i\psi_2(x)). 
\eea
By substituting them into the equations of motion, and 
picking up only linear terms for the fluctuation fields, 
we obtain the linearized equations of motion for the fluctuation fields. 
\bea
 \{\der^m \der_m-\cO_2 \cO_1\}a \eql 0, \nonumber\\
 \{\der^m \der_m-\cO_1 \cO_2\}b \eql 0, \label{ex_EOMb} \\
 i\gm_{(3)}^m \der_m\psi_1-\cO_2\psi_2 \eql 0, \nonumber\\
 i\gm_{(3)}^m \der_m\psi_2-\cO_1\psi_1 \eql 0, \;\;\;\;\;
 \brkt{\der_m\equiv\frac{\der}{\der x^m}} \label{ex_EOMf}
\eea
where 
\bea
 \cO_1 \defa -\der_2-2g\Acl(x_2), \nonumber\\
 \cO_2 \defa \der_2-2g\Acl(x_2). \;\;\;\;\;
 \brkt{\der_2\equiv\frac{\der}{\der x_2}}
\eea
Throughout this paper, the 3D Lorentz indices are denoted by $m$ 
($m=0,1,3$), while the Greek letters~$\mu,\nu,\cdots$ 
are used as the 4D Lorentz indices. 
The matrices~$\gm_{(3)}^m$ are the 3D $\gm$-matrices. 
(See Appendix~\ref{notations}.)

>From Eqs.(\ref{ex_EOMb}) and (\ref{ex_EOMf}), 
we can find the mode equations to be  
\bea
 \cO_1 c_{(n)}(x_2) \eql m_{(n)}d_{(n)}(x_2), \nonumber\\
 \cO_2 d_{(n)}(x_2) \eql m_{(n)}c_{(n)}(x_2), 
 \label{ex_mode_eq}
\eea
where $c_{(n)}(x_2)$ and $d_{(n)}(x_2)$ are the eigenfunctions 
for the eigenvalues $m_{(n)}$, and are called the mode functions. 

By using these mode functions, we can expand the fluctuation fields 
as follows. 
\bea
 A(x) \eql \Acl(x_2)+\frac{1}{\sqrt{2}}\brc{\sum_{n=0}^\infty 
 c_{(n)}(x_2)a_{(n)}(x^m)+i\sum_{n=1}^\infty d_{(n)}(x_2)b_{(n)}(x^m)},
 \nonumber\\
 \Psi^\alpha(x) \eql \frac{1}{\sqrt{2}}\brc{\sum_{n=0}^\infty
 c_{(n)}(x_2)\psi_{1(n)}^\alpha(x^m)
 +i\sum_{n=1}^\infty d_{(n)}(x_2)\psi_{2(n)}^\alpha(x^m)}. 
 \label{mode_expd}
\eea
Note that the operator $\cO_1$ has a zero-mode $c_{(0)}(x_2)$ 
while $\cO_2$ does not. 

Here, $a_{(n)}(x^m)$, $b_{(n)}(x^m)$ and  $\psi_{1(n)}^\alpha(x^m)$, 
$\psi_{2(n)}^\alpha(x^m)$ become 3D real scalar and 
Majorana spinor fields in the resulting 3D effective theory 
with a common mass eigenvalue $m_{(n)}$, respectively. 
Since $a_{(n)}$ and $\psi_{1(n)}^\alpha$, 
or $b_{(n)}$ and $\psi_{2(n)}^\alpha$ have 
a common mode function~$c_{(n)}(x_2)$, 
or $d_{(n)}(x_2)$, they are supposed to form supermultiplets 
for 3D $\cN=1$ SUSY preserved by the wall. 

We can obtain the 3D description of the original theory 
by substituting Eq.(\ref{mode_expd}) into 
the Lagrangian~Eq.(\ref{on_shell_L}) and performing the $x_2$-integration. 


To obtain LET for the zero-modes~$a_{(0)}$ and $\psi_{(0)}$, 
the authors of Ref.\cite{Dvali} simply ignored all massive modes. 
As we will see below, however, this leads to a wrong result. 
The resulting 3D effective Lagrangian $\cL^{(3)}$ is
\be
 \cL^{(3)}=-\frac{1}{2}\der^m a_{(0)}\der_m a_{(0)}
 +\frac{i}{2}\psi_{1(0)}\gm_{(3)}^m \der_m\psi_{1(0)}
 -\frac{\lmd^2}{8}a_{(0)}^4,  \label{nonSUSY_L1}
\ee
where 
\be
 \lmd^2 \equiv 2g^2\int_{-\infty}^\infty \dr x_2 c_{(0)}^4(x_2)
 =\frac{9}{16}g^{5/2}\Lmd. 
\ee

Eq.(\ref{nonSUSY_L1}) is not supersymmetric!
This wrong result stems from the fact that we have eliminated 
the auxiliary field~$F$ before dropping the massive modes. 

For the purpose of understanding the situation, 
let us consider the following three-dimensional model. 
\be
 \cL^{(3)}=\int\dr^2\tht\brc{\frac{1}{2}\cD^\alpha\vph_1 \cD_\alpha\vph_1
 +\frac{1}{2}\cD^\alpha\vph_2 \cD_\alpha\vph_2+m_2\vph_2^2+g\vph_1^2\vph_2},
 \label{ex_model1}
\ee
where 
\be
 \vph_i(x^m,\tht)=a_i(x^m)+\tht\psi_i(x^m)+\frac{1}{2}\tht^2 f_i(x^m) 
 \;\;\;\;\; (i=1,2)
\ee
are 3D scalar superfields, $\tht$ is the 3D Majorana spinor coordinate, 
and $\cD_\alpha$ denotes the covariant derivative for 3D $\cN=1$ SUSY, 
which is defined by Eq.(\ref{3D_cvd}) in Appendix~\ref{cov_derv}. 

In terms of the component fields, Eq.(\ref{ex_model1}) is rewritten 
as follows. 
\bea
 \cL^{(3)} \eql -\frac{1}{2}\der^m a_1\der_m a_1
 -\frac{1}{2}\der^m a_2\der_m a_2+\frac{i}{2}\psi_1\gm_{(3)}^m\der_m\psi_1
 +\frac{i}{2}\psi_2\gm_{(3)}^m\der_m\psi_2 \nonumber\\
 &&+\frac{1}{2}f_1^2+\frac{1}{2}f_2^2+m_2\brkt{a_2 f_2-\frac{1}{2}\psi_2^2}
 \nonumber\\
 &&+g\brc{a_1 f_1 a_2+\frac{1}{2}a_1^2 f_2
 -\frac{1}{2}(2a_1 \psi_1\psi_2+a_2\psi_1^2)}.
\eea
>From the equations of motion for the auxiliary fields, 
\bea
 f_1 \eql -ga_1a_2, \nonumber\\
 f_2 \eql -m_2a_2-\frac{g}{2}a_1^2. 
 \label{f2_exprs}
\eea
Thus, by eliminating the auxiliary fields, we can obtain 
\bea
 \cL^{(3)} \eql -\frac{1}{2}\der^m a_1 \der_m a_1
 -\frac{1}{2}\der^m a_2 \der_m a_2+\frac{i}{2}\psi_1\gm_{(3)}^m\der_m\psi_1
 +\frac{i}{2}\psi_2\gm_{(3)}^m\der_m\psi_2 \nonumber\\
 &&-\frac{1}{2}m_2^2 a_2^2-\frac{1}{2}m_2\psi_2^2
 -ga_1\psi_1\psi_2-\frac{g}{2}a_2\psi_1^2-\frac{1}{2}m_2 g a_1^2 a_2
 -\frac{g^2}{2}a_1^2 a_2^2-\frac{g^2}{8}a_1^4.
 \label{ex_model1_comp}
\eea

Now we will derive LET whose cut-off scale 
is much smaller than the mass scale~$m_2$. 
If we simply ignore the terms that involve massive modes~$a_2$ and $\psi_2$, 
the effective Lagrangian becomes  
\be
 \cL^{(3)}_{\rm eff}=-\frac{1}{2}\der^m a_1 \der_m a_1
 +\frac{i}{2}\psi_1\gm_{(3)}^m \der_m \psi_1-\frac{g^2}{8}a_1^4. 
\ee
Evidently, this is not supersymmetric. 
Of course, this is the consequence of the inadequate procedure 
for integrating out the massive modes. 
The adequate way of integrating out is discussed in Ref.\cite{HLW}. 
To integrate out the heavy modes, they expressed such modes 
as functions of the light modes 
by using the equations of motion for the heavy modes. 
Following their procedure, we can obtain the Lagrangian 
\be
 \cL^{(3)}_{\rm eff}=-\frac{1}{2}\der^m a_1 \der_m a_1
 +\frac{i}{2}\psi_1\gm_{(3)}^m \der_m \psi_1+\cO\brkt{\frac{1}{m_2}}. 
 \label{comp_LET}
\ee
Then, there exists no quartic coupling of $a_1$ at the leading order 
of the $1/m_2$-expansion, and LET certainly  becomes supersymmetric. 

\begin{figure}
\leavevmode
\epsfysize=4cm
\centerline{\epsfbox{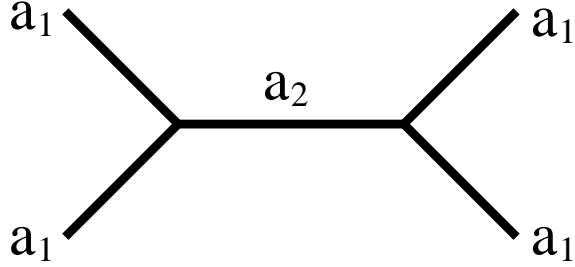}}
\caption{The tree-level diagram that contributes to 
the four-point amplitude~$\cA_{a_1}^{(4)}$. 
}
\label{tree-dgm}
\end{figure}

More explicitly, we can also see that the quartic coupling of $a_1$ vanishes 
at the leading order in the following way. 
Since the decoupling of $a_2$ and $\psi_2$ corresponds to taking the limit 
of $m_2\to\infty$, the cubic coupling of $a_1^2 a_2$ 
in Eq.(\ref{ex_model1_comp}) cannot be neglected. 
Thus a contribution from the tree-diagram in Fig.\ref{tree-dgm} must be 
taken into account. 
Including this contribution, the tree-level 
four-point amplitude~$\cA_{a_1}^{(4)}$ turn out to be zero. 
\be
 \cA_{a_1}^{(4)}=-\frac{g^2}{8}\times 4!
 +\brkt{-\frac{m_2 g}{2}}^2\frac{1}{m_2^2}\times {}_4{\rm C}_2 \times 2
 =0. 
\ee
Here the first term corresponds to the contribution from 
the contact term in the Lagrangian 
and the second term is that from the diagram in Fig.\ref{tree-dgm}. 
The factor~$\frac{1}{m_2^2}$ corresponds to the propagator of 
the massive scalar field~$a_2$, 
and $4!$ and ${}_4{\rm C}_2 \times 2$ are the statistical factors. 
Therefore, no quartic coupling appears in LET. 

The cause of the wrong result~Eq.(\ref{nonSUSY_L1}) is thought to be similar 
to that of the above illustrative model~Eq.(\ref{ex_model1}). 
Namely, we cannot simply ignore the massive modes 
to integrate out them if we have eliminated the auxiliary fields. 

On the other hand, if we keep the superfield description during 
integrating out the massive modes, the situation changes. 
The authors of Ref.\cite{Pomarol} demonstrated such a procedure 
in four dimensions. 
They used the equations of motion for the heavy superfields and 
expressed them as functions of the light ones (and the spurion superfields)
\footnote{
Their purpose was the analysis of the soft SUSY breaking parameters 
appearing in LET of the Grand Unified Theory. 
So they performed the calculation including the soft SUSY breaking terms, 
in contrast to our exact SUSY case. 
}. 
Thus the resulting LET is described by the superfields. 
They showed that the superfield formalism greatly simplifies the calculations. 
For instance, in our illustrative case, we can obtain 
the effective Lagrangian by their procedure, 
\be
 \cL^{(3)}=\int\dr^2\tht\brkt{\frac{1}{2}\cD^\alpha\vph_1
 \cD_\alpha\vph_1}+\cO\brkt{\frac{1}{m_2}}. 
 \label{SUSY_L1}
\ee
This certainly reproduces the previous result~Eq.(\ref{comp_LET}). 
At the leading order, this has the form obtained by simply 
dropping the massive superfield~$\vph_2$ from the original 
Lagrangian~Eq.(\ref{ex_model1}). 
Namely, we can drop the massive modes to integrate out 
if the original theory is described by the superfields. 

Of course, we can obtain the correct LET~Eq.(\ref{comp_LET}) 
from the on-shell expression~Eq.(\ref{ex_model1_comp}) 
if we follow the integrating-out procedure in Ref.\cite{HLW} 
in our illustrative model. 
However, in our original case, such a procedure becomes terribly complicated 
task due to the infinite Kaluza-Klein modes. 
Therefore, in order to obtain the correct LET on the BPS wall, 
we need the 3D superfield description of the original theory, 
which corresponds to 
the expression~Eq.(\ref{ex_model1}) in the above example. 
We will provide such expressions in the rest of the paper.

\section{Embedding 3D superfields into 4D superfield} \label{embedding}
In this section, we will find the way of embedding 3D superfields 
into a 4D chiral superfield. 

The 4D $\cN=1$ SUSY algebra is 
\bea
 \{Q_\alpha,\bar{Q}_{\dot{\beta}}\} \eql 2\sgm^\mu_{\alpha\dot{\beta}}P_\mu, 
 \nonumber\\
 \{Q_\alpha,Q_\beta\} \eql \{\bar{Q}_{\dot{\alpha}},\bar{Q}_{\dot{\beta}}\}=0. 
 \label{4D_SUSYalg}
\eea

The SUSY transformation~$\dlt_\xi$ of 
a chiral supermultiplet~$(A,\Psi^\alpha,F)$ is defined by 
\bea
 \dlt_\xi A \eql \sqrt{2}\xi\Psi, \nonumber\\
 \dlt_\xi \Psi_\alpha \eql i\sqrt{2}(\sgm^\mu\bar{\xi})_\alpha \der_\mu A
 +\sqrt{2}\xi_\alpha F, \nonumber\\
 \dlt_\xi F \eql i\sqrt{2}\bar{\xi}\bar{\sgm}^\mu\der_\mu \Psi. 
 \label{SUSYtrf}
\eea

Define actions of the generators $P_\mu$, $Q_\alpha$ 
and $\bar{Q}^{\dot{\alpha}}$ on the fields $\phi=A,\Psi^\alpha,F$ as 
\bea
 P_\mu \times \phi=-i\der_\mu\phi, \nonumber\\
 (\xi Q+\bar{\xi}\bar{Q})\times \phi=\dlt_\xi \phi. 
\eea
Then, we can check that the SUSY transformation~Eq.(\ref{SUSYtrf}) 
is certainly a representation of the SUSY algebra~Eq.(\ref{4D_SUSYalg}). 

Notice that a chiral superfield $\Phi$ can be written as 
\be
 \Phi(x,\tht,\btht)=e^{\dlt_\tht}\times A(x)
 =\brkt{1+\dlt_\tht+\frac{1}{2}\dlt_\tht^2}\times A(x). 
\ee
Then, we can express $\Phi$ as 
\be
 \Phi(x,\tht,\btht)=\Omega\times A(0),
\ee
where 
\be
 \Omega\equiv e^{ix^\mu P_\mu+\tht Q+\btht\bar{Q}}. 
 \label{def_Omega}
\ee

Upon the chiral superfield~$\Phi$, the generators $P_\mu$, $Q_\alpha$ 
and $\bar{Q}^{\dot{\alpha}}$ 
can be represented by the following differential operators. 
\bea
 \hat{P}_\mu \eql -i\der_\mu, \nonumber\\
 \hat{Q}_\alpha \eql \der_\alpha-i(\sgm^\mu\btht)_\alpha\der_\mu, 
 \nonumber\\
 \hat{\bar{Q}}^{\dot{\alpha}} \eql 
 \der^{\dot{\alpha}}-i(\bar{\sgm}^\mu\tht)^{\dot{\alpha}}\der_\mu. 
\eea
where
\be
 \der_\alpha\equiv\frac{\der}{\der\tht^\alpha}, \;\;\;
 \der^{\dot{\alpha}}\equiv\frac{\der}{\der\btht_{\dot{\alpha}}} 
\ee
are left-derivatives. 

Here, we will rewrite Eq.(\ref{SUSYtrf}) to the form which is 
convenient for the following discussion. 
First, we redefine spinors as $\eta^\alpha\to\eta^\alpha$, 
$\eta_\alpha\to -i\eta_\alpha$, 
so that $\eta_\alpha=(\sgm^2)_{\alpha\beta}\eta^\beta$. 
Next, we express $\sgm^\mu$ and $\bar{\sgm}^\mu$ in terms of 
the 3D $\gm$-matrices $\gm_{(3)}^m$ 
through Eq.(\ref{sgm_gm3}) in Appendix~\ref{notations}. 
Furthermore, we decompose the transformation parameter~$\xi$ as follows. 
\be
 \xi^\alpha=\frac{e^{i\dlt/2}}{\sqrt{2}}(\xi_1^\alpha+i\xi_2^\alpha), 
 \label{xi_decomp}
\ee
where $\xi_i^\alpha$ ($i=1,2$) are 3D Majorana spinors 
($(\xi_i^\alpha)^*=\xi_i^\alpha$), 
and $\dlt$ is a phase determined by the wall configuration. 
(See Eq.(\ref{choice_dlt}).)
As a result, Eq.(\ref{SUSYtrf}) is rewritten as 
\bea
 \dlt_\xi A \eql -ie^{i\dlt/2}\xi_1\Psi+e^{i\dlt/2}\xi_2\Psi, \nonumber\\
 \dlt_\xi \Psi_\alpha \eql e^{-i\dlt/2}\brc{
 (\gm_{(3)}^m\xi_1)_\alpha\der_m A-\xi_{1\alpha}\der_2 A
 +e^{i\dlt}\xi_{1\alpha}F} \nonumber\\
 &&-ie^{-i\dlt/2}\brc{(\gm_{(3)}^m\xi_2)_\alpha\der_m A
 -\xi_{2\alpha}\der_2 A-e^{i\dlt}\xi_{2\alpha}F}, \nonumber\\
 \dlt_\xi F \eql -ie^{-i\dlt/2}\brc{\xi_1\gm_{(3)}^m\der_m\Psi
 +\xi_1\der_2\Psi}
 -e^{-i\dlt/2}\brc{\xi_2\gm_{(3)}^m\der_m\Psi+\xi_2\der_2\Psi}. 
 \label{decomp_SUSYtrf}
\eea

Corresponding to the decomposition Eq.(\ref{xi_decomp}), 
we also decompose the 4D supercharges as follows. 
\be
 Q_\alpha=\frac{e^{-i\dlt/2}}{\sqrt{2}}(Q_{1\alpha}-iQ_{2\alpha}), 
 \;\;\;
 \bar{Q}_{\dot{\alpha}}=-(Q_\alpha)^*
 =-\frac{e^{i\dlt/2}}{\sqrt{2}}(Q_{1\alpha}+iQ_{2\alpha}), 
 \label{4D_Q_decomp}
\ee
where $(Q_{i\alpha})^*=Q_{i\alpha}$ and $(Q_i^\alpha)^*=-Q_i^\alpha$. 
Then it follows that 
\be
 \tht Q+\btht\bar{Q}=\tht_1 Q_1+\tht_2 Q_2. 
\ee
By the above definition of $Q_1$ and $Q_2$, 
$Q_1$ becomes the broken supercharge 
and $Q_2$ is the unbroken supercharge by the wall. 

Under the decomposition Eq.(\ref{4D_Q_decomp}), 
the SUSY algebra Eq.(\ref{4D_SUSYalg}) becomes 
\bea
 \{Q_{1\alpha},Q_{1\beta}\} \eql \{Q_{2\alpha},Q_{2\beta}\}
 =2(\gm_{(3)}^m \sgm^2)_{\alpha\beta}P_m, \nonumber\\
 \{Q_{1\alpha},Q_{2\beta}\} \eql -\{Q_{2\alpha},Q_{1\beta}\}
 =2i(\sgm^2)_{\alpha\beta}P_2. 
 \label{3D_SUSYalg}
\eea
This can be interpreted as the (central extended) 3D $\cN=2$ SUSY algebra 
if we identify $P_2$ with the central charge. 

Hence, $\Omega$ defined by Eq.(\ref{def_Omega}) can be rewritten as 
\be
 \Omega=e^{ix^mP_m+ix_2P_2+\tht_1Q_1+\tht_2Q_2}
 =e^{ix^mP_m+i(x_2-\tht_1\tht_2)P_2+\tht_2Q_2}e^{\tht_1Q_1}. 
\ee
Now, define a group element, 
\be
 \tl{\Omega}\equiv e^{ix^mP_m+ix_2P_2+\tht_2Q_2}e^{\tht_1Q_1}, 
\ee
then we can show that 
\bea
 Q_{2\alpha}\tl{\Omega} \eql \brc{\der_{2\alpha}
 +i(\gm_{(3)}^m\tht_2)_\alpha\der_m}\tl{\Omega}, \nonumber\\
 Q_{1\alpha}\tl{\Omega} \eql \brc{\der_{1\alpha}
 +i(\gm_{(3)}^m\tht_1)_\alpha\der_m-2\tht_{2\alpha}\der_2}\tl{\Omega}, 
\eea
where 
\be
 \der_{1\alpha}\equiv\frac{\der}{\der\tht_1^\alpha}, \;\;\;
 \der_{2\alpha}\equiv\frac{\der}{\der\tht_2^\alpha}, 
\ee
are left-derivatives. 

Therefore, if we define  
\be
 \tl{\Phi}(x^m,x_2,\tht_1,\tht_2)\equiv 
 \Phi(x^m,x_2+\tht_1\tht_2,\tht_1,\tht_2)=\tl{\Omega}\times A(0), 
 \label{tl_Phi}
\ee
the representation of each generator on $\tl{\Phi}$ is 
\bea
 \hat{P}_m \eql -i\der_m, \nonumber\\
 \hat{P}_2 \eql -i\der_2, \nonumber\\
 \hat{Q}_{1\alpha} \eql \der_{1\alpha}+i(\gm_{(3)}^m\tht_1)_\alpha\der_m 
 -2\tht_{2\alpha}\der_2, \nonumber\\
 \hat{Q}_{2\alpha} \eql \der_{2\alpha}+i(\gm_{(3)}^m\tht_2)_\alpha\der_m. 
\eea
Namely, the unbroken SUSY $Q_2$ is represented by the usual form 
of 3D $\cN=1$ superspace on $\tl{\Phi}$. 
(See Eq.(\ref{hatQ3}) in Appendix~\ref{cov_derv}.) 

Then, we will decompose $\tl{\Phi}$ into 3D superfields. 
>From Eq.(\ref{decomp_SUSYtrf}), 
\be
 Q_{1\alpha}\times A=-ie^{i\dlt/2}\Psi_\alpha=-iQ_{2\alpha}\times A. 
\ee
Using this relation, we can convert $Q_1$ acting on $A$ 
into $Q_2$ and $P_2$. 
\be
 e^{\tht_1Q_1}\times A=e^{-i\tht_1Q_2-\tht_1^2P_2}\times A. 
 \label{Q1toQ2}
\ee

Thus, 
\bea
 \tl{\Phi}=\tl{\Omega}\times A(0) \eql e^{ix^mP_m+ix_2P_2+\tht_2Q_2}
 e^{-i\tht_1Q_2-\tht_1^2P_2}\times A(0) \nonumber\\
 \eql e^{-i\tht_1D_2+i\tht_1^2\der_2}e^{ix^mP_m+ix_2P_2+\tht_2Q_2}
 \times A(0).  \label{expr_tl_Omg}
\eea
Here we have used the formula~Eq.(\ref{DW_WG_ex}) 
in Appendix~\ref{cov_derv}. 

Therefore, if we introduce a quantity
\be
 \vph(x^m,x_2,\tht_2)\equiv e^{ix^mP_m+ix_2P_2+\tht_2 Q_2}\times A(0). 
 \label{def_vph}
\ee
the following relation can be obtained. 
\be
 \Phi(x^m,x_2+\tht_1\tht_2,\tht_1,\tht_2)
 =e^{-i\tht_1D_2+i\tht_1^2\der_2}\vph(x^m,x_2,\tht_2). 
 \label{tl_Phi-vph}
\ee

Note that $\vph$ behaves like a 3D scalar superfield under $Q_2$-SUSY, 
though the component fields are still four-dimensional fields. 

In fact, if we expand $\vph$ in terms of $\tht_2$ as 
\be
 \vph=a+\tht_2\psi+\frac{1}{2}\tht_2^2 f, 
\ee
the transformation of the component fields is read off as 
\bea
 \dlt_{\xi_2}a \eql \xi_2 \psi, \nonumber\\
 \dlt_{\xi_2}\psi_\alpha \eql -i(\gm_{(3)}^m\xi_2)_\alpha\der_m a
 +\xi_{2\alpha}f, \nonumber\\
 \dlt_{\xi_2}f \eql -i\xi_2\gm_{(3)}^m\der_m\psi. 
 \label{3D_SUSYtrf}
\eea

By noticing $a=A$ and comparing Eq.(\ref{decomp_SUSYtrf}) 
and Eq.(\ref{3D_SUSYtrf}), the relations of the component fields of $\vph$ 
to the original fields are\footnote{
Note the redefinition of spinors mentioned above Eq.(\ref{xi_decomp}). 
}
\bea
 a \eql A, \nonumber\\
 \psi^\alpha \eql e^{i\dlt/2}\Psi^\alpha, \nonumber\\
 f \eql i(\der_2 A+e^{i\dlt}F). \label{rel_a-A}
\eea

The result of this section is Eq.(\ref{tl_Phi-vph}). 
If $\vph$ is mode-expanded, Eq.(\ref{tl_Phi-vph}) will provide the relation 
between the 4D chiral superfield and the 3D scalar superfields. 
In order to carry out the mode-expansion, we need the mode-equation 
for $\vph$. 
We will derive it in the next section.

\section{Derivation of 3D effective theory} \label{3D_LET_deriv}
In this section, we will expand 
the 4D chiral superfield in terms of 3D superfields,  
and derive 3D effective theory, 
which is manifestly supersymmetric, 
by executing the $x_2$- and $\tht_1$-integrations. 

Here we will consider the following generalized Wess-Zumino model 
as a four-dimensional bulk theory. 
\bea
 \cL \eql \int\dr^2\tht{\rm d}^2\btht\; K(\bar{\Phi},\Phi)
 +\int\dr^2\tht\; W(\Phi)+\int\dr^2\btht\; \bar{W}(\bar{\Phi}) \nonumber\\
 \eql K_{i\bar{j}}\brc{F^i\bar{F}^{\bar{j}}
 -\der^\mu A^i\der_\mu\bar{A}^{\bar{j}}
 -i\bar{\Psi}^{\bar{j}}\bar{\sgm}^\mu D_\mu\Psi^i} \nonumber\\
 &&-\frac{1}{2}\Psi^i\Psi^jK_{l\bar{k}}\Gm^l_{ij}\bar{F}^{\bar{k}} 
 -\frac{1}{2}\bar{\Psi}^{\bar{i}}\bar{\Psi}^{\bar{j}}K_{k\bar{l}}
 \Gm^{\bar{l}}_{\bar{i}\bar{j}}F^k
 +\frac{1}{4}K_{ij\bar{k}\bar{l}}\Psi^i\Psi^j\bar{\Psi}^{\bar{k}}
 \bar{\Psi}^{\bar{l}} \nonumber\\
 &&+F^i W_i-\frac{1}{2}W_{ij}\Psi^i\Psi^j
 +\bar{F}^{\bar{i}}\bar{W}_{\bar{i}}
 -\frac{1}{2}\bar{W}_{\bar{i}\bar{j}}\bar{\Psi}^{\bar{i}}\bar{\Psi}^{\bar{j}}. 
 \label{GWZ}
\eea
Lower indices denote derivatives in terms of corresponding chiral 
or anti-chiral superfields. 
For instance, 
\be
 K_{i\bar{j}}\equiv\frac{\der^2 K}{\der\Phi^i\der\bar{\Phi}^{\bar{j}}}. 
\ee
$\Gm^l_{ij}$ and $\Gm^{\bar{l}}_{\bar{i}\bar{j}}$ are 
the connections on the K\"{a}hler manifold 
and defined by 
\be
 \Gm^l_{ij}\equiv K^{\bar{k}l}K_{ij\bar{k}}, \;\;\;\;\;
 \Gm^{\bar{l}}_{\bar{i}\bar{j}}\equiv K^{\bar{l}k}K_{\bar{i}\bar{j}k}, 
\ee
where $K^{\bar{k}l}$ is the inverse matrix of 
the K\"{a}hler metric~$K_{l\bar{k}}$. 
The definitions of integral measures~${\rm d}\tht$ and ${\rm d}\btht$ 
are listed in Appendix~\ref{notations}.

\subsection{Equation of motion}
The equation of motion of the above theory 
can be expressed by the superfields as follows. 
\be
 -\frac{1}{4}\bar{D}^2K_i+W_i=0.  \label{sfEOM}
\ee

We will express this equation of motion in terms of 
$\vph(x^m,x_2,\tht_2)$ defined by Eq.(\ref{def_vph}).

For the second term in L.H.S. of Eq.(\ref{sfEOM}), such rewriting 
can easily be done. 
Note that $W_i(\Phi)$ is a chiral superfield. 
Then, by repeating the procedure in the previous section 
with the replacement of $A$ with $W_i(A)$, 
we can obtain 
\be
 W_i(\Phi)=e^{-\tht_1\tht_2\der_2}e^{-i\tht_1D_2+i\tht_1^2\der_2}W_i(\vph). 
\ee

On the other hand, rewriting the first term in L.H.S. of Eq.(\ref{sfEOM}) 
is somewhat complicated. 

Using the formula~Eq.(\ref{DW_WG}) in Appendix~\ref{cov_derv}, 
\be
 \bar{D}^2\bar{\Phi}=\bar{D}^2(\Omega\times\bar{A})
 =\Omega\times(\bar{Q}^2\times\bar{A}). \label{D2_bPhi}
\ee
Since  
\be
 \bar{Q}^2=e^{i\dlt}\brc{Q_2 Q_1+2iP_2-\frac{i}{2}(Q_1^2-Q_2^2)} 
\ee
from Eq.(\ref{4D_Q_decomp}) and Eq.(\ref{3D_SUSYalg}), 
we can calculate $\bar{Q}^2\times\bar{A}$ as  
\be
 \bar{Q}^2\times\bar{A}=2ie^{i\dlt}(Q_2^2+2P_2)\times\bar{A}. 
 \label{bQ2bA}
\ee
Here we have used the complex conjugate of Eq.(\ref{Q1toQ2}). 

Since $\bar{D}^2\bar{\Phi}$ is a chiral superfield, 
it can be written in the similar form of Eq.(\ref{tl_Phi-vph}). 
In this case, the quantity corresponding to $\vph$ is 
\be
 \chi\equiv e^{ix^mP_m+ix_2P_2+\tht_2Q_2}\times
 \brkt{\bar{Q}^2\times\bar{A}(0)}=2ie^{i\dlt}(D_2^2-2i\der_2)\bar{\vph}. 
\ee
In the second equation, we have used Eq.(\ref{bQ2bA}) 
and the formula~Eq.(\ref{DW_WG_ex}). 

Then, Eq.(\ref{D2_bPhi}) can be written as 
\bea
 \bar{D}^2\bar{\Phi} \eql e^{-i\tht_1\tht_2P_2}\tl{\Omega}\times 
 (\bar{Q}^2\times\bar{A})
 =e^{-\tht_1\tht_2\der_2}e^{-i\tht_1D_2+i\tht_1^2\der_2}\chi 
 \nonumber\\
 \eql 2ie^{i\dlt}e^{-\tht_1\tht_2\der_2}e^{-i\tht_1D_2+i\tht_1^2\der_2}
 (D_2^2-2i\der_2)\bar{\vph}. 
 \label{bDsbPhi}
\eea

Since a product of two chiral superfields 
$\Phi^1=e^{-\tht_1\tht_2\der_2}e^{-i\tht_1D_2+i\tht_1^2\der_2}\vph^1$ 
and $\Phi^2=e^{-\tht_1\tht_2\der_2}e^{-i\tht_1D_2+i\tht_1^2\der_2}\vph^2$ is 
also a chiral superfield, we can easily show that 
\be
 \Phi^1\Phi^2=e^{-\tht_1\tht_2\der_2}e^{-i\tht_1D_2+i\tht_1^2\der_2}
 \vph^1\vph^2.
\ee
Thus, for 
\be
 K_i=\sum\kappa_{i_1\cdots i_n\bar{j}_1\cdots\bar{j}_m}
 \Phi^{i_1}\cdots\Phi^{i_n}\bar{\Phi}^{\bar{j}_1}
 \cdots\bar{\Phi}^{\bar{j}_m}, 
\ee
we can obtain the expression 
\bea
 \bar{D}^2K_i \eql \sum\kappa_{i_1\cdots i_n\bar{j}_1\cdots\bar{j}_m}
 \Phi^{i_1}\cdots\Phi^{i_n}\bar{D}^2\brkt{\bar{\Phi}^{\bar{j}_1}
 \cdots\bar{\Phi}^{\bar{j}_m}} \nonumber\\
 \eql 2ie^{i\dlt}e^{-\tht_1\tht_2\der_2}e^{-i\tht_1D_2+i\tht_1^2\der_2}
 \sum\kappa_{i_1\cdots i_n\bar{j}_1\cdots\bar{j}_m}
 \brkt{\vph^{i_1}\cdots\vph^{i_n}(D_2^2-2i\der_2)
 (\bar{\vph}^{\bar{j}_1}\cdots\bar{\vph}^{\bar{j}_m})} \nonumber\\
 \eql 2ie^{i\dlt}e^{-\tht_1\tht_2\der_2}e^{-i\tht_1D_2+i\tht_1^2\der_2}
 \brkt{D_2^\alpha(K_{i\bar{j}}D_{2\alpha}\bar{\vph}^{\bar{j}})
 -K_{ik\bar{j}}D_2^\alpha\vph^kD_{2\alpha}\bar{\vph}^{\bar{j}}
 -2iK_{i\bar{j}}\der_2\bar{\vph}^{\bar{j}}}. \nonumber\\ 
 \eql 2ie^{i\dlt}e^{-\tht_1\tht_2\der_2}e^{-i\tht_1D_2+i\tht_1^2\der_2}
 \brkt{K_{i\bar{j}}D_2^2\bar{\vph}^{\bar{j}}
 +K_{i\bar{j}\bar{k}}D_2^\alpha\bar{\vph}^{\bar{j}}
 D_{2\alpha}\bar{\vph}^{\bar{k}}
 -2iK_{i\bar{j}}\der_2\bar{\vph}^{\bar{j}}}. \nonumber\\ 
 \label{bD2Ki}
\eea
In the second equation, we have used Eq.(\ref{bDsbPhi}). 

As a result, the equations of motion Eq.(\ref{sfEOM}) can be rewritten as 
\be
 -\frac{i}{2}\brkt{K_{i\bar{j}}D_2^2\bar{\vph}^{\bar{j}}
 +K_{i\bar{j}\bar{k}}D_2^\alpha\bar{\vph}^{\bar{j}}
 D_{2\alpha}\bar{\vph}^{\bar{k}}}
 -K_{i\bar{j}}\der_2\bar{\vph}^{\bar{j}}+e^{-i\dlt}W_i=0. 
 \label{vphEOM}
\ee

\subsection{BPS equation}
The BPS equation for the domain wall can be obtained 
from the minimal energy condition for the classical field configuration. 
The energy (per unit area) of the domain wall is expressed by 
\bea
 E \eql \int\dr x_2 \brc{K_{i\bar{j}}\der_2 A^i\der_2\bar{A}^{\bar{j}}
 +K^{\bar{j}i}W_i\bar{W}_{\bar{j}}} \nonumber\\
 \eql \int\dr x_2 \brc{K_{i\bar{j}}
 (\der_2 A^i-e^{i\dlt}K^{\bar{k}i}\bar{W}_{\bar{k}})
 (\der_2\bar{A}^{\bar{j}}-e^{-i\dlt}K^{\bar{j}l}W_l)
 +e^{i\dlt}\bar{W}_{\bar{j}}\der_2\bar{A}^{\bar{j}}
 +e^{-i\dlt}W_i\der_2 A^i} \nonumber\\
 \!\!\!&\geq\!\!\!& 
 \int\dr x_2 \brc{e^{i\dlt}\der_2\bar{W}+e^{-i\dlt}\der_2 W}
 =2\int\dr x_2 \der_2\Re(e^{-i\dlt}W)
 =2\Re(e^{-i\dlt}\Dlt W), 
 \label{E_bound}
\eea
where 
\be
 \Dlt W\equiv \int_{\Gm}{\rm d}W
\ee
 and $\Gm$ is the orbit 
for the classical field configuration 
on the target space of the scalar fields. 

In the case that the extra dimension (the $x_2$-direction) is non-compact, 
$\Dlt W$ depends only on the values of the superpotential at the end points 
of $\Gm$, that is, 
\be
 \Dlt W=W(x_2=\infty)-W(x_2=-\infty). 
\ee 

On the other hand, when the extra dimension is compactified on $S^1$, 
$\Gm$ must be a non-contractible cycle in order for the field configuration 
to be topologically stable. 
Furthermore, $W$ must be a multi-valued function\footnote{
Of course, ${\rm d}W$ must be a single-valued function 
since it determines the scalar potential. 
} 
because non-zero $\Dlt W$ is needed for the existence of 
the BPS field configuration\cite{HLS}. 
In this case, the value of $\Dlt W$ is determined 
by a homotopy class where $\Gm$ belongs.

The equality in Eq.(\ref{E_bound}) holds when the scalar fields satisfy 
the equation
\be
 \der_2 A^i=e^{i\dlt}K^{\bar{j}i}\bar{W}_{\bar{j}}. 
 \label{BPSeq} 
\ee 

>From Eq.(\ref{E_bound}), the most stringent bound is 
\be
 E\geq 2|\Dlt W|, 
\ee
which comes from the case that the phase~$\dlt$ is chosen as 
\be
 \dlt=\arg(\Dlt W). \label{choice_dlt}
\ee 
Eq.(\ref{BPSeq}) with this choice of $\dlt$ is called 
{\it the BPS equation}. 

When the classical field configuration~$\Acl^i(x_2)$ satisfies 
Eq.(\ref{BPSeq}), 
we can see that a half of the original supersymmetry, {\it i.e.} 
$Q_2$-SUSY, is preserved  
from Eq.(\ref{decomp_SUSYtrf}), 
\be
 \dlt_{\xi_2}\phi=0 \;\;\;\;\;
 (\phi=A,\Psi,F)
\ee
Here we have used the equation of motion for 
the auxiliary fields~$F^i$, 
\be
 F^i=\frac{1}{2}\Gm^i_{jk}\Psi^j\Psi^k-K^{i\bar{j}}\bar{W}_{\bar{j}}, 
\ee
and $\Psi_{\rm cl}^i=0$. 

For a solution of Eq.(\ref{BPSeq}) $\Acl^i(x_2)$, 
let us define a quantity 
\be
 \vpcl^i\equiv e^{ix^mP_m+ix_2P_2+\tht_2Q_2}\times \Acl^i(0). 
\ee

Then, $\vpcl^i$ becomes a solution of the equations of motion 
Eq.(\ref{vphEOM}). 
In fact, since $\psi_{\rm cl}^i=0$ and $f_{\rm cl}^i=0$ 
from Eq.(\ref{rel_a-A}) and the BPS equations, we can see  
\be
 \vpcl^i=\Acl^i(x_2). 
\ee
Thus, 
\be
 D_{2\alpha}\bar{\vph}_{\rm cl}^{\bar{i}}=0, 
\ee
and from Eq.(\ref{BPSeq}),
\be
 K_{i\bar{j}}\der_2\bar{\vph}_{\rm cl}^{\bar{j}}
 -e^{-i\dlt}W_i(\vpcl)=0.  \label{clBPSeq}
\ee
Therefore, $\vpcl^i$ certainly satisfy the equations of 
motion~Eq.(\ref{vphEOM}).

\subsection{Mode expansion of the fluctuation fields}
Next, we will consider the equations of motion for 
the fluctuation fields~$\tl{\vph}^i$ 
around the classical solution~$\vpcl^i$. 
Substituting $\vph^i=\vpcl^i+\tl{\vph}^i$ into Eq.(\ref{vphEOM}), 
we obtain 
\bea
 &&-\frac{i}{2}K_{i\bar{j}}(\vpcl)D_2^2\bar{\tl{\vph}}^{\bar{j}}
 -\brc{K_{i\bar{j}k}(\vpcl)\tl{\vph}^k+K_{i\bar{j}\bar{k}}(\vpcl)
 \bar{\tl{\vph}}^{\bar{k}}}\der_2\bar{\vpcl}^{\bar{j}} \nonumber\\ 
 &&\hspace{3.5cm}-K_{i\bar{j}}(\vpcl)\der_2\bar{\tl{\vph}}^{\bar{j}}
 +e^{-i\dlt}W_{ij}(\vpcl)\tl{\vph}^j+\cdots =0,  
\eea
where ellipsis denotes the higher order terms for $\tl{\vph}$ or 
$\bar{\tl{\vph}}$. 

Using Eq.(\ref{clBPSeq}), the equations of motion become 
\be
 -\frac{1}{2}K_{i\bar{j}}(\vpcl)D_2^2\bar{\tl{\vph}}^{\bar{j}}
 +i\brc{\cD_y \bar{\tl{\vph}}_i-e^{-i\dlt}\cD_i W_j \tl{\vph}^j}
 +\cdots =0, \label{vphEOM2}
\ee
where 
\be
 \bar{\tl{\vph}}_i\equiv K_{i\bar{j}}\bar{\tl{\vph}}^{\bar{j}}, \;\;\;
 \cD_y\bar{\tl{\vph}}_i\equiv \der_2\bar{\tl{\vph}}_i
 -\Gm^k_{ij}\der_2\vpcl^j\bar{\tl{\vph}}_k, \;\;\;
 \cD_i W_j\equiv W_{ij}-\Gm^k_{ij}W_k. 
\ee

>From Eq.(\ref{vphEOM2}), we can find the mode equation,  
\be
 i\brc{\cD_y\bar{u}_{(n)i}-e^{-i\dlt}\cD_i W_j u^j_{(n)}}
 =m_{(n)}\bar{u}_{(n)i},  \label{mode_eq}
\ee
The eigenfunctions of this equation~$\bar{u}_{(n)i}$ are called 
the mode functions. 

Now we expand $\tl{\vph}^i$ by the mode functions~$u_{(n)}^i(x_2)$ 
($n=0,1,2,\cdots$). 
\be
 \tl{\vph}^i(x^m,x_2,\tht_2)=\frac{1}{\sqrt{2}}\sum_{n=0}^\infty 
 u_{(n)}^i(x_2)\vph_{(n)}(x^m,\tht_2). 
 \label{mode_expand}
\ee

As we will see in the following, $\vph_{(n)}$ becomes 
a 3D $\cN=1$ scalar superfield with the mass~$m_{(n)}$. 
In particular, Eq.(\ref{mode_eq}) has a zero-mode 
$u_{(0)}^i(x_2)=C\der_2\Acl^i(x_2)$ ($C$ is a real normalization factor), 
which corresponds to the Nambu-Goldstone mode for the translational 
invariance and $Q_1$-SUSY. 

When the eigenvalues of Eq.(\ref{mode_eq}) $m_{(n)}$ are all real, 
we can show the orthogonal relation of the mode functions~$u_{(n)}^i(x_2)$, 
(See Appendix~\ref{orthogonal}.) 
\be
 \Re\brc{\int\dr x_2 \bar{u}_{(n)i}(x_2)u_{(m)}^i(x_2)}=\dlt_{nm}. 
 \label{orthonorm}
\ee
In the following, the mode functions are supposed to be normalized.

\subsection{3D effective theory}
Now we will express the original theory Eq.(\ref{GWZ}) in terms of 
$\vph^i$ and $\bar{\vph}^{\bar{j}}$ defined by Eq.(\ref{def_vph}), 
and carry out the $\tht_1$- and 
$x_2$-integration in order to obtain LET on the wall.

First, we will express the K\"{a}hler potential term in terms of 
$\vph^i$ and $\bar{\vph}^{\bar{j}}$. 

Under the $x$-integration, note that 
\be
 \int\dr^2\tht{\rm d}^2\btht\; K(\bar{\Phi},\Phi)
 =\int\dr^2\tht \brkt{-\frac{1}{4}\bar{D}^2 K}. 
\ee
By the same procedure as that for the derivation of Eq.(\ref{bD2Ki}), 
we can show that 
\be
 \bar{D}^2 K=2ie^{i\dlt}e^{-\tht_1\tht_2\der_2}
 e^{-i\tht_1D_2+i\tht_1^2\der_2}
 \brc{D_2^\alpha(K_{\bar{j}}D_{2\alpha}\bar{\vph}^{\bar{j}})
 -K_{i\bar{j}}D_2^\alpha\vph^i D_{2\alpha}\bar{\vph}^{\bar{j}}
 -2iK_{\bar{j}}\der_2\bar{\vph}^{\bar{j}}}. 
\ee
Thus, up to the total derivatives, we can obtain the expression 
\bea
 \int\dr^2\tht{\rm d}^2\btht\; K(\bar{\Phi},\Phi) \eql \nonumber\\
 &&\hspace{-2.5cm}
 \int\dr^2\tht \left[-\frac{i}{2}e^{i\dlt}e^{-i\tht_1D_2}
 \brc{D_2^\alpha(K_{\bar{j}}D_{2\alpha}\bar{\vph}^{\bar{j}})
 -K_{i\bar{j}}D_2^\alpha\vph^iD_{2\alpha}\bar{\vph}^{\bar{j}}
 -2iK_{\bar{j}}\der_2\bar{\vph}^{\bar{j}}}\right]. 
 \label{intK}
\eea

Noting that (see Appendix~\ref{notations})
\be
 {\rm d}^2\tht{\rm d}^2\btht=-{\rm d}^2\tht_1{\rm d}^2\tht_2, \;\;\;\;\;\;
 \bar{\tht}^2=e^{-i\dlt}\brc{\tht_1\tht_2+\frac{i}{2}(\tht_1^2-\tht_2^2)}, 
\ee
it follows that 
\bea
 \int\dr^2\tht\; e^{-i\tht_1D_2}
 \eql \int\dr^2\tht{\rm d}^2\btht \;\btht^2 e^{-i\tht_1D_2}
 \nonumber\\ 
 \eql -\int\dr^2\tht_1{\rm d}^2\tht_2 \;e^{-i\dlt}
 \brc{\tht_1\tht_2+\frac{i}{2}(\tht_1^2-\tht_2^2)}
 e^{-i\tht_1D_2} \nonumber\\
 \eql -\int\dr^2\tht_2 \;\frac{i}{2}e^{-i\dlt}e^{\tht_2D_2}. 
\eea
The explicit appearance of $\tht_2$ in the integrand seems to break 
$Q_2$-SUSY at first sight, but it can be absorbed into 
$\vph$ and $\bar{\vph}$ as follows. 

Under the $x$-integration, 
it can be shown for an arbitrary function~$\cF$ that 
\bea
 \int\dr^2\tht\;e^{-i\tht_1D_2}\cF(\vph,\bar{\vph})
 \eql \int\dr^2\tht_2
 \brc{-\frac{i}{2}e^{-i\dlt}e^{\tht_2D_2}\cF(\vph,\bar{\vph})}
 \nonumber\\
 \eql \int\dr^2\tht_2
 \brc{-\frac{i}{2}e^{-i\dlt}e^{\tht_2D_2}e^{ix^mP_m+ix_2P_2+\tht_2Q_2}\times
 \cF(a(0),\bar{a}(0))} \nonumber\\
 \eql \int\dr^2\tht_2
 \brc{-\frac{i}{2}e^{-i\dlt}e^{ix^mP_m+ix_2P_2+2\tht_2Q_2}\times
 \cF(a(0),\bar{a}(0))} \nonumber\\
 \eql \int\dr^2\tht_2 \brc{-2ie^{-i\dlt}\cF(\vph,\bar{\vph})}. 
 \label{formula1}
\eea
Here we have used the formula~Eq.(\ref{DW_WG_ex}) in the third step, 
and changed the integration variable $2\tht_2\to\tht_2$ in the last step. 

Using this formula, Eq.(\ref{intK}) can be expressed as 
\bea
 \int\dr^2\tht{\rm d}^2\btht\; K(\bar{\Phi},\Phi) 
 \eql \int\dr^2\tht_2
 \left[-\brc{D_2^\alpha(K_{\bar{j}}D_{2\alpha}\bar{\vph}^{\bar{j}})
 -K_{i\bar{j}}D_2^\alpha\vph^iD_{2\alpha}\bar{\vph}^{\bar{j}}
 -2iK_{\bar{j}}\der_2\bar{\vph}^{\bar{j}}}\right] \nonumber\\
 \eql \int\dr^2\tht_2
 \brc{K_{i\bar{j}}D_2^\alpha\vph^iD_{2\alpha}\bar{\vph}^{\bar{j}}
 -2iK_i\der_2\vph^i}. 
\eea
Here we have dropped the total derivatives. 

The superpotential terms can easily be rewritten in terms of $\vph^i$ 
by using the formula~Eq.(\ref{formula1}), 
\bea
 \int\dr^2\tht \;W(\Phi)
 \eql \int\dr^2\tht\; e^{-\tht_1\tht_2\der_2}
 e^{-i\tht_1D_2+i\tht_1^2\der_2}W(\vph) \nonumber\\
 \eql \int\dr^2\tht_2 \brc{-2ie^{-i\dlt}W(\vph)}. 
\eea
The total derivatives have been dropped again. 

As a result, the action after the $\tht_1$-integration is 
\be
 S=\int\dr^3 x\int\dr^2\tht_2\int\dr x_2
 \brc{K_{i\bar{j}}D_2^\alpha\vph^i D_{2\alpha}\bar{\vph}^{\bar{j}}
 -2iK_i\der_2\vph^i+4\Im\brkt{e^{-i\dlt}W(\vph)}}. 
 \label{tht1intS}
\ee

Substituting $\vph=\vpcl+\tl{\vph}$ into Eq.(\ref{tht1intS}), 
the effective Lagrangian becomes 
\bea
 \cL^{(3)} \eql \int\dr^2\tht_2\int\dr x_2
 \left\{K_{i\bar{j}}(\vpcl)D_2^\alpha\tl{\vph}^iD_{2\alpha}
 \bar{\tl{\vph}}^{\bar{j}}\right. \nonumber\\
 &&-i\bar{\tl{\vph}}^{\bar{i}}\brkt{
 K_{\bar{i}jk}(\vpcl)\der_2\vpcl^j\tl{\vph}^k
 +K_{\bar{i}j\bar{k}}(\vpcl)\der_2\vpcl^j\bar{\tl{\vph}}^{\bar{k}}
 +K_{j\bar{i}}(\vpcl)\der_2\tl{\vph}^j
 -e^{i\dlt}\bar{W}_{\bar{i}\bar{j}}(\bar{\vph}_{\rm cl})
 \bar{\tl{\vph}}^{\bar{j}}} \nonumber\\
 &&+i\tl{\vph}^i\brkt{
 K_{i\bar{j}k}(\vpcl)\der_2\bar{\vph}_{\rm cl}^{\bar{j}}\tl{\vph}^k
 +K_{i\bar{j}\bar{k}}(\vpcl)\der_2\bar{\vph}_{\rm cl}^{\bar{j}}
 \bar{\tl{\vph}}^{\bar{k}}
 +K_{i\bar{j}}(\vpcl)\der_2\bar{\tl{\vph}}^{\bar{j}}
 -e^{-i\dlt}W_{ij}(\vpcl)\tl{\vph}^j} \nonumber\\
 &&+\cdots\}, 
\eea
where the ellipsis denotes the higher terms for $\tl{\vph}$ 
or $\bar{\tl{\vph}}$. 

Then, we will expand $\tl{\vph}$ as Eq.(\ref{mode_expand}), 
and use the mode equations~Eq.(\ref{mode_eq}) and 
the orthonormalization of the mode functions~Eq.(\ref{orthonorm}), 
so that we can obtain the desired 3D effective Lagrangian. 
\bea
 \cL^{(3)} \eql \int\dr^2\tht_2 \left[
 \sum_{n=0}^\infty\brc{\frac{1}{2}(D_2\vph_{(n)})^2+m_{(n)}\vph_{(n)}^2}
 +\sum_{m,n,l}g_{m(nl)}\vph_{(m)}D_2\vph_{(n)}D_2\vph_{(l)}\right. 
 \nonumber\\
 &&\hspace{1.2cm}
 \left.-\sum_{m,n,l}\lmd_{(mnl)}\vph_{(m)}\vph_{(n)}\vph_{(l)}+\cdots\right], 
\eea
where the complete symmetrization is supposed 
for indices in the parentheses, and 
\be
 g_{mnl}=\frac{1}{2\sqrt{2}}\int\dr x_2\brc{
 K_{i\bar{j}k}(\vpcl)u_{(m)}^k u_{(n)}^i \bar{u}_{(l)}^{\bar{j}}
 +K_{i\bar{j}\bar{k}}(\vpcl)\bar{u}_{(m)}^{\bar{k}}u_{(n)}^i
 \bar{u}_{(l)}^{\bar{j}}}, 
 \label{g_mnl}
\ee
\bea
 \lmd_{mnl}\eql \frac{1}{2\sqrt{2}}\int\dr x_2 \;
 \Im\left\{K_{ijk}(\vpcl)\der_2 u_{(m)}^i u_{(n)}^j u_{(l)}^k
 +2K_{ij\bar{k}}(\vpcl)\der_2 u_{(m)}^i u_{(n)}^j \bar{u}_{(l)}^{\bar{k}}
 \right.\nonumber\\
 &&\hspace{2.5cm}
 +K_{i\bar{j}\bar{k}}(\vpcl)\der_2 u_{(m)}^i \bar{u}_{(n)}^{\bar{j}}
 \bar{u}_{(l)}^{\bar{k}}
 +\frac{1}{3}K_{ijkh}(\vpcl)\der_2\vpcl^i u_{(m)}^j u_{(n)}^k u_{(l)}^h
 \nonumber\\
 &&\hspace{2.5cm}
 +K_{ijk\bar{h}}(\vpcl)\der_2\vpcl^i u_{(m)}^j u_{(n)}^k 
 \bar{u}_{(l)}^{\bar{h}}
 +K_{ij\bar{k}\bar{h}}(\vpcl)\der_2\vpcl^i u_{(m)}^j \bar{u}_{(n)}^{\bar{k}}
 \bar{u}_{(l)}^{\bar{h}} \nonumber\\
 &&\hspace{2.5cm}
 \left.+\frac{1}{3}K_{i\bar{j}\bar{k}\bar{h}}(\vpcl)\der_2\vpcl^i
 \bar{u}_{(m)}^{\bar{j}}\bar{u}_{(n)}^{\bar{k}}\bar{u}_{(l)}^{\bar{h}}
 +\frac{2}{3}e^{-i\dlt}W_{ijk}(\vpcl)u_{(m)}^i u_{(n)}^j u_{(l)}^k\right\}. 
 \nonumber\\
 \label{lmd_mnl}
\eea

\subsection{Case of real wall-configurations} \label{real_config}
Before concluding this section, we will apply the above result to 
a simple case, where 
the 4D bulk theory involves only one chiral superfield, 
the K\"{a}hler potential is minimal, and all the parameters in the theory 
and the classical field configuration~$\Acl(x_2)$ are real. 

In this case, 
the mode equation~Eq.(\ref{mode_eq}) can be written as 
\be
 i\brc{\der_2\bar{u}_{(n)}-\frac{\der^2 W}{\der\Phi^2}(\Acl)u_{(n)}}
 =m_{(n)}\bar{u}_{(n)}. 
\ee
Taking the complex conjugation, we can obtain 
\be
 i\brc{\der_2 u_{(n)}-\frac{\der^2 W}{\der\Phi^2}(\Acl)\bar{u}_{(n)}}
 =-m_{(n)}u_{(n)}. 
\ee
These equations mean that when $u_{(n)}$ is the mode function 
with the eigenvalue~$m_{(n)}$, $\bar{u}_{(n)}$ is also the mode function 
whose eigenvalue is $-m_{(n)}$. 
Namely, the mass spectrum in this case is doubly degenerate. 
(As we will see, the zero-mode is exceptional.)
Considering this fact, the mode-expansion of $\tl{\vph}$ is 
\be
 \tl{\vph}=\frac{1}{\sqrt{2}}\sum_{n=0}^\infty 
 \brc{u_{(n)}\vph_{(+n)}+\bar{u}_{(n)}\vph_{(-n)}}. 
 \label{mode_exp1}
\ee
Signs in the label of the mode functions denote those of 
the corresponding mass eigenvalues. 

Here we decompose each mode function into the real and imaginary parts. 
\be
 u_{(n)}=\frac{1}{\sqrt{2}}(u_{{\rm R}(n)}+iu_{{\rm I}(n)}). 
 \label{u_decomp}
\ee
Then, the mode expansion~Eq.(\ref{mode_exp1}) becomes 
\be
 \tl{\vph}=\frac{1}{\sqrt{2}}\sum_{n=0}^\infty
 \brc{u_{{\rm R}(n)}\vph_{{\rm R}(n)}+iu_{{\rm I}(n)}\vph_{{\rm I}(n)}},
\ee
where 
\be
 \vct{\vph_{{\rm I}(n)}}{\vph_{{\rm R}(n)}}
 =\frac{1}{\sqrt{2}}\mtrx{1}{-1}{1}{1}\vct{\vph_{(+n)}}{\vph_{(-n)}}. 
 \label{rot_vph}
\ee
Therefore we can see that the decomposition~Eq.(\ref{u_decomp}) corresponds to 
the rotation between the degenerate modes. 

In the basis of $\vph_{{\rm R}(n)}$ and $\vph_{{\rm I}(n)}$, 
the mode equation becomes the following combined equations. 
\bea
 \brkt{-\der_2+\frac{\der^2 W}{\der\Phi^2}}u_{{\rm R}(n)}
 \eql m_{(n)}u_{{\rm I}(n)}, \nonumber\\
 \brkt{\der_2+\frac{\der^2 W}{\der\Phi^2}}u_{{\rm I}(n)}
 \eql m_{(n)}u_{{\rm R}(n)}. 
 \label{comb_md_eq}
\eea
These correspond to Eq.(\ref{ex_mode_eq}) in Section~\ref{motivation}. 
>From these equations, we can see that the zero-mode is non-degenerate. 
In fact, the zero-mode exists only in the first equation 
of Eq.(\ref{comb_md_eq}).\footnote{ 
The zero-mode solution of the second equation of Eq.(\ref{comb_md_eq}) 
diverges at $x_2\to\pm\infty$. 
}
Namely, 
\be
 u_{(0)}=\frac{1}{\sqrt{2}}u_{{\rm R}(0)}=\bar{u}_{(0)}. 
\ee
Therefore, Eq.(\ref{mode_exp1}) can be rewritten as 
\be
 \tl{\vph}=\frac{1}{\sqrt{2}}\sum_{n=-\infty}^\infty 
 u_{(n)}\vph_{(n)}, 
\ee
where each mode function satisfies the following constraint 
\be
 u_{(-n)}=\bar{u}_{(n)}. 
\ee
Then, the effective Lagrangian~$\cL^{(3)}$ is written as 
\bea
 \cL^{(3)} \eql \int\dr^2\tht_2 \left[
 \sum_{n=-\infty}^\infty \brc{\frac{1}{2}(D_2\vph_{(n)})^2
 +m_{(n)}\vph_{(n)}^2}+\cdots \right] \nonumber\\
 \eql \int\dr^2\tht_2 \left[\frac{1}{2}(D_2\vph_{(0)})^2+
 \sum_{n=1}^\infty\left\{\frac{1}{2}(D_2\vph_{(n)})^2
 +\frac{1}{2}(D_2\vph_{(-n)})^2 \right.\right. \nonumber\\
 &&\hspace{2cm}\left.\left.+(\vph_{(n)},\vph_{(-n)})
 \mtrx{m_{(n)}}{}{}{-m_{(n)}}
 \vct{\vph_{(n)}}{\vph_{(-n)}}\right\}+\cdots\right]. 
\eea

In the basis of $\vph_{{\rm R}(n)}$ and $\vph_{{\rm I}(n)}$, 
it can be rewritten as 
\bea
 \cL^{(3)}  \eql \int\dr^2\tht_2 \left[\frac{1}{2}(D_2\vph_{(0)})^2+
 \sum_{n=1}^\infty\left\{\frac{1}{2}(D_2\vph_{{\rm R}(n)})^2
 +\frac{1}{2}(D_2\vph_{{\rm I}(n)})^2  \right.\right.\nonumber\\
 &&\hspace{2cm}\left.\left.+(\vph_{{\rm I}(n)},\vph_{{\rm R}(n)})
 \mtrx{}{m_{(n)}}{m_{(n)}}{}
 \vct{\vph_{{\rm I}(n)}}{\vph_{{\rm R}(n)}}\right\}+\cdots\right]. 
\eea
Up to the quadratic terms, this expression coincides with the form derived in 
Ref.\cite{Dvali,Chibisov} after eliminating the auxiliary fields. 

The cubic couplings in Eqs.(\ref{g_mnl}) and (\ref{lmd_mnl}) are 
simplified in this case. 
\bea
 g_{mnl} \eql g_{m(nl)}=0, \nonumber\\ 
 \lmd_{mnl} \eql \lmd_{(mnl)}=\frac{1}{3\sqrt{2}}\int\dr x_2\;
 \Im\brc{\frac{\der^3 W}{\der\Phi^3}(\Acl)
 u_{(n)}u_{(m)}u_{(l)}}. 
\eea
Since the zero-mode function~$u_{(0)}$ is real, 
we can see the zero mode~$\vph_{(0)}$ does not have 
cubic self-interactions. 
This is true for all higher order couplings. 
This result coincides with that of Ref.\cite{Chibisov}. 

However, this is not the case when $\Acl(x_2)$ is a complex field 
configuration.

\section{Dimensional reduction of the gauge theories}
\label{vector_SF} 
So far, we have discussed the derivation of 3D LET 
from 4D theory that contains only chiral superfields. 
In this section, we will derive 3D LET including a vector supermultiplet. 
Here, we will suppose that the gauge supermultiplet 
does not couple to the scalar fields 
that contribute to the wall configuration. 
So the gauge symmetry is unbroken by the wall, and 
the zero-mode of the 4D gauge field, 
which corresponds to the 3D gauge field in LET, lives in the bulk. 
Therefore, the extra dimension of the wall ($x_2$-direction) 
is supposed to be compactified on $S^1$ in this section\footnote{
BPS domain walls in such a case are discussed in Ref.\cite{HLS}. 
}.
Here we will concentrate ourselves to the case of the abelian gauge 
supermultiplet, for simplicity. 

\subsection{Supertransformation of the gauge supermultiplet}
The 4D vector superfield~$V(x,\tht,\btht)$ is written by 
\bea
 V(x,\tht,\btht) \eql C+i\tht\chi-i\tht\bar{\chi}
 +i\tht^2B-i\btht^2\bar{B}-\tht\sgm^\mu\btht v_\mu \nonumber\\
 &&+i\tht^2\btht^2\brkt{\bar{\lmd}+\frac{i}{2}\bar{\sgm}^\mu\der_\mu\chi}
 -i\btht^2\tht\brkt{\lmd+\frac{i}{2}\sgm^\mu\der_\mu\bar{\chi}}
 +\frac{1}{2}\tht^2\btht^2\brkt{D+\frac{1}{2}\der^\mu\der_\mu C}, 
 \nonumber\\ \label{vct_sf}
\eea
where $C$ and $D$ are real scalars, $\chi$ and $\lmd$ are complex 
Weyl spinors, and $B$ is a complex scalar. 

The SUSY transformations of the component fields are 
\bea
 \dlt_\xi C \eql i(\xi\chi-\bar{\xi}\bar{\chi}), \nonumber\\
 \dlt_\xi \chi_\alpha \eql (\sgm^\mu\bar{\xi})_\alpha 
 (\der_\mu C+iv_\mu)+2\xi B, \nonumber\\
 \dlt_\xi B \eql \bar{\xi}\bar{\lmd}+i\bar{\xi}\bar{\sgm}^\mu\der_\mu\chi, 
 \nonumber\\
 \dlt_\xi v_\mu \eql i\xi\sgm_\mu\bar{\lmd}+i\bar{\xi}\bar{\sgm}_\mu\lmd
 +\xi\der_\mu\chi+\bar{\xi}\der_\mu\bar{\chi}, \nonumber\\
 \dlt_\xi \lmd_\alpha \eql i\xi_\alpha D+(\sgm^{\mu\nu}\xi)_\alpha v_{\mu\nu}, 
 \nonumber\\
 \dlt_\xi D \eql -\xi\sgm^\mu\der_\mu\bar{\lmd}
 +\bar{\xi}\bar{\sgm}^\mu\der_\mu\lmd, 
 \label{4vectSUSYtrf}
\eea
where $v_{\mu\nu}\equiv \der_\mu v_\nu-\der_\nu v_\mu$ is the field strength. 

Now, we will again redefine spinors and rewrite $\sgm$-, 
$\bar{\sgm}$-matrices in terms of 
the 3D $\gm$-matrices $\gm_{(3)}^m$, 
as we did around Eq.(\ref{xi_decomp}). 
Furthermore, the following decompositions are performed. 
\bea
 \xi^\alpha \eql \frac{e^{i\dlt/2}}{\sqrt{2}}(\xi_1^\alpha+i\xi_2^\alpha), 
 \nonumber\\
 \chi^\alpha \eql \frac{e^{-i\dlt/2}}{\sqrt{2}}(\chi_1^\alpha+i\chi_2^\alpha), 
 \nonumber\\
 \lmd^\alpha \eql \frac{e^{i\dlt/2}}{\sqrt{2}}(\lmd_1^\alpha+i\lmd_2^\alpha), 
 \nonumber\\
 B \eql \frac{e^{-i\dlt}}{2}(M+iN), 
\eea
where $\dlt$ is a phase defined in Eq.(\ref{choice_dlt}). 
Then, Eq.(\ref{4vectSUSYtrf}) is rewritten as 
\bea
 \dlt_\xi C \eql \xi_1\chi_1-\xi_2\chi_2, \nonumber\\
 \dlt_\xi \chi_{1\alpha} \eql -i(\gm_{(3)}^m\xi_1)_\alpha\der_m C
 -\xi_{1\alpha}v_2+\xi_{1\alpha}M
 -i(\gm_{(3)}^m\xi_2)_\alpha v_m+\xi_{2\alpha}\der_2 C-\xi_{2\alpha}N, 
 \nonumber\\
 \dlt_\xi \chi_{2\alpha} \eql -i(\gm_{(3)}^m\xi_1)_\alpha v_m
 +\xi_{1\alpha}\der_2 C+\xi_{1\alpha}N
 +i(\gm_{(3)}^m\xi_2)_\alpha\der_m C+\xi_{2\alpha}v_2+\xi_{2\alpha}M, 
 \nonumber\\
 \dlt_\xi M \eql \xi_1\lmd_2-i\xi_1\gm_{(3)}^m\der_m\chi_1
 +\xi_1\der_2\chi_2+\xi_2\lmd_1-i\xi_2\gm_{(3)}^m\der_m\chi_2
 -\xi_2\der_2\chi_1, \nonumber\\
 \dlt_\xi N \eql \xi_1\lmd_1-i\xi_1\gm_{(3)}^m\der_m\chi_2
 -\xi_1\der_2\chi_1-\xi_2\lmd_2+i\xi_2\gm_{(3)}^m\der_m\chi_1
 -\xi_2\der_2\chi_2, \nonumber\\
 \dlt_\xi v_m \eql -i\xi_1\gm_{(3)m}\lmd_1+\xi_1\der_m\chi_2
 -i\xi_2\gm_{(3)m}\lmd_2+\xi_2\der_m\chi_1, \nonumber\\
 \dlt_\xi v_2 \eql \xi_1\lmd_2+\xi_1\der_2\chi_2-\xi_2\lmd_1
 +\xi_2\der_2\chi_1, \nonumber\\
 \dlt_\xi \lmd_{1\alpha} \eql (\gm_{(3)}^{mn}\xi_1)_\alpha v_{mn}
 -\xi_{2\alpha}D+i(\gm_{(3)}^m\xi_2)_\alpha v_{m2}, \nonumber\\
 \dlt_\xi \lmd_{2\alpha} \eql \xi_{1\alpha}D-i(\gm_{(3)}^m\xi_1)_\alpha 
 v_{m2}+(\gm_{(3)}^{mn}\xi_2)_\alpha v_{mn}, \nonumber\\
 \dlt_\xi D \eql -i\xi_1\gm_{(3)}^m\der_m\lmd_2-\xi_1\der_2\lmd_1
 +i\xi_2\gm_{(3)}^m\der_m\lmd_1-\xi_2\der_2\lmd_2. 
 \label{vctSUSYtrf_decomp}
\eea

Note that the vector superfield~Eq.(\ref{vct_sf}) can be expressed as 
\be
 V=e^{ix^\mu P_\mu+\tht Q+\btht\bar{Q}}\times C(0)
 =e^{ix^m P_m+i(x_2-\tht_1\tht_2)P_2+\tht_2 Q_2}e^{\tht_1 Q_1}
 \times C(0). 
\ee
Using Eq.(\ref{vctSUSYtrf_decomp}), 
\be
 e^{\tht_1 Q_1}\times C=C+\tht_1\chi_1
 +\frac{1}{2}\tht_1^2(-v_2+M). 
\ee
Then, if we introduce the following quantities, 
\bea
 \kappa(x^m,x_2,\tht_2) \defa e^{ix^m P_m+ix_2 P_2+\tht_2 Q_2}
 \times C(0), \nonumber\\
 \rho_\alpha(x^m,x_2,\tht_2) \defa e^{ix^m P_m+ix_2 P_2+\tht_2 Q_2}
 \times \chi_{1\alpha}(0), \nonumber\\
 \sgm(x^m,x_2,\tht_2) \defa e^{ix^m P_m+ix_2 P_2+\tht_2 Q_2}
 \times \frac{1}{2}(-v_2(0)+M(0)), 
 \label{def_krs}
\eea
we can express $V$ as\footnote{
This $\tht_1$-expansion is analogous to the $\cN=1$ decomposition 
of the 3D $\cN=2$ vector superfield presented in Ref.\cite{Zupnik}. 
} 
\be
 V(x^m,x_2+\tht_1\tht_2,\tht_1,\tht_2)=
 \kappa(x^m,x^2,\tht_2)+\tht_1\rho(x^m,x^2,\tht_2)
 +\tht_1^2\sgm(x^m,x^2,\tht_2). \label{V_tht1_expd}
\ee

\subsection{Gauge transformation and gauge fixing}
Next, we will discuss the gauge transformation. 
The gauge transformation for the vector superfield is 
\be
 V\to V+\Lmd+\bar{\Lmd}, \label{4Dgauge_trf}
\ee
where the transformation parameter~$\Lmd$ is a chiral superfield, 
{\it i.e.}~$\bar{D}_{\dot{\alpha}}\Lmd=0$. 
>From Eq.(\ref{tl_Phi-vph}), $\Lmd$ is expressed in the following form.  
\bea
 \Lmd(x^m,x_2+\tht_1\tht_2,\tht_1,\tht_2) \eql 
 e^{-i\tht_1 D_2+i\tht_1^2\der_2}\beta(x^m,x_2,\tht_2) \nonumber\\
 \eql \beta-i\tht_1D_2\beta+\tht_1^2\brkt{
 \frac{1}{4}D_2^2\beta+i\der_2\beta}, 
\eea
where $\beta$ is a quantity defined by 
$\beta\equiv e^{ix^m P_m+ix_2 P_2+\tht_2 Q_2}\times 
\Lmd(x^\mu=\tht=\btht=0)$. 
Then, the transformation~Eq.(\ref{4Dgauge_trf}) is rewritten in terms of 
the quantities defined in Eq.(\ref{def_krs}) as 
\bea
 \kappa & \to & \kappa+\beta+\bar{\beta}, \nonumber\\
 \rho_\alpha & \to & \rho_\alpha-iD_{2\alpha}(\beta-\bar{\beta}), \nonumber\\
 \sgm & \to & \sgm+\frac{1}{4}D_2^2(\beta+\bar{\beta})
 +i\der_2(\beta-\bar{\beta}). 
\eea

Using this gauge transformation, we can greatly simplify the expression 
of the vector superfield. 
First, $\kappa$ can be set to zero by choosing the gauge parameter $\beta$ as 
\be
 2\Re\beta=-\kappa. \label{Rebeta}
\ee
This means that $\kappa$ is a pure gauge degree of freedom. 
On the other hand, $\rho_\alpha$ and $\sgm$ correspond to a 3D 
$\cN=1$ vector and scalar superfield components of 
a 3D $\cN=2$ vector supermultiplet. 

After eliminating $\kappa$ by the gauge transformation, there is still 
a gauge degree of freedom corresponding to the gauge transformation 
parameter~$\Im\beta$.
>From Eqs.(\ref{vctSUSYtrf_decomp}) and (\ref{def_krs}), 
$\rho_\alpha$ is expanded as  
\be
 \rho_\alpha=\chi_{1\alpha}-i(\gm_{(3)}^m\tht_2)_\alpha v_m
 +\tht_{2\alpha}(\der_2 C-N)+\tht_2^2
 \brkt{-\lmd_{2\alpha}+\frac{i}{2}(\gm_{(3)}^m\der_m\chi_1)_\alpha}. 
\ee
Then, by choosing the components of the gauge parameter 
$2\Im\beta=a+\tht_2\psi+\frac{1}{2}\tht_2^2 f$ as 
\bea
 \psi^\alpha \eql -\chi_1^\alpha, \nonumber\\
 f \eql -(\der_2 C-N), \label{Imbeta}
\eea
we can eliminate the auxiliary fields $\chi_1^\alpha$, $C$ and $N$, 
and obtain a simple expression, 
\be
 \rho_\alpha=-i(\gm_{(3)}^m\tht_2)_\alpha v_m^{\prime}-\tht_2^2\lmd_{2\alpha},
\ee
where
\be
 v_m^{\prime}\equiv v_m+\der_m a. \label{3Dgaugetrf_cpnt}
\ee
Here, Eq.(\ref{3Dgaugetrf_cpnt}) represents the usual 3D gauge transformation. 
In the choice of the gauge parameter~Eqs.(\ref{Rebeta}) and (\ref{Imbeta}), 
the expression of $\sgm$ becomes 
\bea
 \sgm \eql e^{\tht_2 Q_2}\times \frac{1}{2}(-v_2+M)
 -\frac{1}{4}D_2^2\brkt{e^{\tht_2Q_2}\times C}-\der_2(2\Im\beta)
 \nonumber\\
 \eql e^{\tht_2Q_2}\times(-v_2)-\der_2(2\Im\beta) \nonumber\\
 \eql -(v_2+\der_2 a)+\tht_2\lmd_1-\frac{1}{2}\tht_2^2 D. 
 \label{gf_sgm}
\eea

Namely, all the 4D auxiliary fields of $V$ is eliminated 
in this gauge. 
Hence, the gauge choice~Eqs.(\ref{Rebeta}) and (\ref{Imbeta}) 
corresponds to the Wess-Zumino gauge in four dimensions. 
After these gauge fixings, there are still residual gauge degrees of freedom, 
which associate with the parameter~$a(x)$. 
At first sight, it seems that the 3D scalar~$v_2$ 
can completely be eliminated by choosing the gauge parameter~$a$ so that 
\be
 \der_2 a=-v_2. \label{a_gf1}
\ee
However, it must be noticed that $a(x)$ does not contain the mode that is 
linear to $x_2$ because of the periodicity of $a(x)$ 
along the $x_2$-direction, {\it i.e.} 
\be
 a(x^m,x_2+2\pi R)=a(x^m,x_2). 
\ee
Therefore the zero-mode~$v_{2(0)}$ cannot be gauged away by the gauge fixing, 
and is a {\it physical} mode. 
So Eq.(\ref{a_gf1}) should be modified as  
\be
 \der_2 a=-v_2|_{\rm nz}, \label{a_gf2}
\ee
where $v_2|_{\rm nz}\equiv v_2-v_{2(0)}$ contains only non-zero modes. 

Note that the solution of Eq.(\ref{a_gf2}) has an ambiguity of $a_{(0)}(x^m)$, 
which is independent of $x_2$. 
This means that there is a residual gauge symmetry 
whose transformation parameter is $a_{(0)}(x^m)$. 
This is the 3D gauge symmetry in LET. 

In the following, we will take the gauge mentioned above. 
In this gauge, Eq.(\ref{V_tht1_expd}) is simplified as 
\be
 V(x^m,x_2+\tht_1\tht_2,\tht_1,\tht_2)
 =\tht_1\rho(x^m,x_2,\tht_2)+\tht_1^2\sgm(x^m,x_2,\tht_2), 
\ee
where 
\bea
 \rho_\alpha \eql -i(\gm_{(3)}^m\tht_2)_\alpha v_m-\tht_2^2\lmd_{2\alpha}, 
 \nonumber\\
 \sgm \eql -v_{2(0)}+\tht_2\lmd_1-\frac{1}{2}\tht_2^2 D. 
\eea

The gauge invariant quantities are 
\bea
 w_\alpha \defa \frac{1}{4}D_2^2\rho_\alpha
 +\frac{i}{2}(\gm_{(3)}^m\der_m\rho)_\alpha
 =e^{\tht_2Q_2}\times \lmd_{2\alpha}, \nonumber\\
 u_\alpha \defa D_{2\alpha}\sgm+\der_2\rho_\alpha
 =e^{\tht_2Q_2}\times \lmd_{1\alpha}. 
 \label{def_w_u}
\eea
Here, $w_\alpha$ is the 3D superfield strength. 
Eq.(\ref{def_w_u}) is written in terms of the component fields as 
\bea
 w_\alpha \eql \lmd_{2\alpha}+(\gm_{(3)}^{mn}\tht_2)_\alpha v_{mn}
 -\frac{i}{2}\tht_2^2(\gm_{(3)}^m\der_m\lmd_2)_\alpha, \nonumber\\
 u_\alpha \eql \lmd_{1\alpha}-\tht_{2\alpha}D+i(\gm_{(3)}^m\tht_2)_\alpha
 (\der_m v_{2(0)}-\der_2 v_m)+\tht_2^2\brc{\frac{i}{2}
 (\gm_{(3)}^m\der_m\lmd_1)_\alpha-\der_2\lmd_{2\alpha}}. 
\eea

These quantities are related to the 4D superfield 
strength~$W_\alpha\equiv -\frac{1}{4}\bar{D}^2D_\alpha V$ through 
\be
 W_\alpha(x^m,x_2+\tht_1\tht_2,\tht_1,\tht_2)
 =-\frac{i}{\sqrt{2}}e^{i\dlt/2}e^{-i\tht_1D_2+i\tht_1^2\der_2}
 \brc{u_\alpha(x^m,x_2,\tht_2)+iw_\alpha(x^m,x_2,\tht_2)}. 
 \label{rel_W-wu}
\ee
Since $W_\alpha$ is a chiral superfield, 
this relation can also be obtained by the procedure discussed 
in the previous section.

\subsection{3D effective theory}
Now we will derive 3D LET 
by carrying out the integration in terms of $\tht_1$ and $x_2$.

The gauge kinetic term of the 4D theory is 
\be
 \cL_{\rm gauge}=\frac{1}{4}\int\dr^2\tht f(\Phi)W^\alpha W_\alpha
 +{\rm h.c.}, \label{Lgauge_kin}
\ee
where $f(\Phi)$ is a holomorphic function of $\Phi$, 
called the gauge kinetic function, 
and $f(0)=1$ from the requirement that $W_\alpha$ is canonically 
normalized. 

Substituting Eqs.(\ref{tl_Phi-vph}) and (\ref{rel_W-wu}) 
into Eq.(\ref{Lgauge_kin}) and 
following a similar procedure to that in the previous section, 
we can obtain the $\tht_1$-integrated expression, 
\bea
 \cL_{\rm gauge} \eql \int\dr^2\tht_2\left[
 \frac{\Re f(\vph)}{2}\brc{(D_2\sgm)^2+2D_2^\alpha\sgm\der_2\rho_\alpha
 +(\der_2\rho)^2-w^2}\right. \nonumber\\
 &&\hspace{1.4cm}\left.
 -\Im f(\vph)\brc{(D_2^\alpha\sgm+\der_2\rho^\alpha)w_\alpha}\right]. 
 \label{tht1intS2}
\eea

Since the gauge multiplet does not feel the existence of the domain wall 
at the classical level, the mode expansion of it is trivial, that is, 
\bea
 \rho^\alpha(x^m,x_2,\tht_2) \eql \frac{1}{\sqrt{2\pi R}}
 \rho_{(0)}^\alpha(x^m,\tht_2) \nonumber\\
 &&+\sum_{n=1}^\infty\frac{1}{\sqrt{\pi R}}\brc{
 \cos\frac{nx_2}{R}\cdot\rho_{(n+)}^\alpha(x^m,\tht_2)
 +\sin\frac{nx_2}{R}\cdot\rho_{(n-)}^\alpha(x^m,\tht_2)}, \nonumber\\
 \sgm(x^m,x_2,\tht_2) \eql \frac{1}{\sqrt{2\pi R}}
 \sgm_{(0)}(x^m,\tht_2) \nonumber\\
 &&+\sum_{n=1}^\infty\frac{1}{\sqrt{\pi R}}\brc{
 \cos\frac{nx_2}{R}\cdot\sgm_{(n+)}(x^m,\tht_2)
 +\sin\frac{nx_2}{R}\cdot\sgm_{(n-)}(x^m,\tht_2)},  \nonumber\\ 
 \label{md_expd_V}
\eea
where signs in the label of the 3D superfields denote the parity charge 
of the mode functions under $x_2\to -x_2$. 
Thus, by carrying out the $x_2$-integration, we can obtain 
the following effective Lagrangian. 
\bea
 \cL^{(3)}_{\rm gauge} \eql \int\dr^2\tht_2 \left[\frac{1}{2}(D_2\sgm_{(0)})^2
 -\frac{1}{2}w_{(0)}^2 \right.  \nonumber\\
 &&\hspace{1.3cm}+\frac{1}{2}\sum_{n=1}^\infty\left\{
 (D_2\sgm_{(n+)})^2+(D_2\sgm_{(n-)})^2+2\frac{n}{R}
 (D_2\sgm_{(n+)}\rho_{(n-)}-D_2\sgm_{(n-)}\rho_{(n+)})\right. \nonumber\\
 &&\hspace{2.8cm}\left.\left.+\frac{n^2}{R^2}(\rho_{(n+)}^2+\rho_{(n-)}^2)
 -w_{(n+)}^2-w_{(n-)}^2\right\}+\cdots\right] 
 \label{vctEFT}
\eea
Here we used the assumption that scalar fields that 
couple to the gauge field do not have nontrivial background configurations, 
that is, $f(\Phi_{\rm cl})=f(0)=1$. 

The first two terms in Eq.(\ref{vctEFT}) represent the kinetic terms 
of the 3D $\cN=2$ gauge multiplet, 
and the quadratic terms in the second and the third lines 
correspond to the Kaluza-Klein modes with masses $m_{\rm K.K.}=n/R$. 

Since we have already eliminated the non-zero modes of $v_2$, 
terms such as $\der_m v_{2(n\pm)}v^m_{(n\mp)}$ are absent.

\subsection{Couplings to the matter}
Finally, we will write down the gauge couplings to the matter 
supermultiplet. 
For simplicity, we will consider the case that 
the K\"{a}hler potential is minimal and the chiral matter superfield~$\Phi$ 
does not have a non-trivial classical configuration, {\it i.e.} 
$\Phi_{\rm cl}=0$. 

Here, note that the abelian gauge symmetry should be represented as 
$O(2)$ symmetry in our case since 3D superfields are real. 
Thus, the vector superfield~$V$ discussed so far should be understood as 
a $2\times 2$ matrix 
\be
 V=V_R\mtrx{0}{-i}{i}{0}, 
\ee
where $V_R$ is a real vector supermultiplet, 
and the matter superfield~$\Phi$ should be understood as a 2-component 
column vector whose gauge transformation is 
\be
 \Phi\to\exp\brc{-2g\Lmd\mtrx{0}{-i}{i}{0}}\Phi, 
\ee 
where $g$ is a gauge coupling constant. 

Then, the 4D gauge coupling is written by 
\bea
 \cL_{\rm matter} \eql \int\dr^2\tht{\rm d}^2\btht\; \bar{\Phi}e^{2gV}\Phi
 \nonumber\\
 \eql -\int\dr^2\tht_1{\rm d}^2\tht_2
 \brc{\bar{\vph}+i\tht_1D_2\bar{\vph}+\tht_1^2\brkt{
 \frac{1}{4}D_2^2\bar{\vph}-i\der_2\bar{\vph}}}
 \brc{1+2g\tht_1\rho+\tht_1^2(2g\sgm-g^2\rho^2)} \nonumber\\
 &&\hspace{2.2cm}\times\brc{\vph-i\tht_1D_2\vph+\tht_1^2\brkt{
 \frac{1}{4}D_2^2\vph+i\der_2\vph}} \nonumber\\
 \eql \int\dr^2\tht_2 \left[D_2^\alpha\bar{\vph}D_{2\alpha}\vph
 +2\Im\brkt{\bar{\vph}\der_2\vph}
 +2g\Im\brkt{\bar{\vph}\rho D_2\vph}
 -\bar{\vph}(2g\sgm-g^2\rho^2)\vph \right], 
 \nonumber\\ \label{tht1intS3}
\eea
where $\vph$ is a 2-component column vector, and $\rho_\alpha$ and $\sgm$ 
are $2\times 2$ matrices. 
We can obtain the effective theory by expanding each superfield 
into the Kaluza-Klein modes and performing the $x_2$-integration. 
\bea
 \cL_{\rm matter}^{(3)} \eql 
 \int\dr^2\tht_2\left[\sum_{n=0}^\infty \left\{
 \frac{1}{2}(D_2\vph_{(n)})^2+m_{(n)}\vph_{(n)}^2
 +g_{(0)}\Im\brkt{{}^t \vph_{(n)}\rho_{(0)}D_2\vph_{(n)}}
 +\frac{1}{2}g_{(0)}^2{}^t\vph_{(n)}\rho_{(0)}^2\vph_{(n)}
 \right\}\right. \nonumber\\
 &&\hspace{1.4cm}+\sum_{n,m=0}^\infty \sum_{l=1}^\infty \sum_{s=\pm}
 g_{R(nm,ls)}\brc{\Im\brkt{{}^t\vph_{(n)}\rho_{(ls)}D_2\vph_{(m)}}
 +g_{(0)}{}^t\vph_{(n)}\rho_{(ls)}\rho_{(0)}\vph_{(m)}} \nonumber\\
 &&\hspace{1.4cm}-\sum_{n,m=0}^\infty \brc{g_{I(nm,0)}{}^t\vph_{(n)}
 \sgm_{(0)}\vph_{(m)}+\sum_{l=1}^\infty\sum_{s=\pm}g_{I(nm,ls)}
 {}^t\vph_{(n)}\sgm_{(ls)}\vph_{(m)}} \nonumber\\
 &&\hspace{1.4cm}\left.+\frac{1}{2}\sum_{n,m=0}^\infty\sum_{l,p=1}^\infty
 \sum_{s,t=\pm}
 (g^2)_{(nm,ls,pt)}{}^t\vph_{(n)}\rho_{(ls)}\rho_{(pt)}\vph_{(m)}
 \right], 
 \label{minimal_g_int}
\eea
where $g_{(0)}\equiv g/\sqrt{2\pi R}$ is the three-dimensional 
gauge coupling, and the other effective couplings are defined as follows. 
\bea
 g_{I(nm,0)} \defa \frac{g}{\sqrt{2\pi R}}\int\dr x_2\;
 \Im\brkt{\bar{u}_{(n)}(x_2)u_{(m)}(x_2)}, \nonumber\\
 g_{R(nm,l+)} \defa \frac{g}{\sqrt{\pi R}}\int\dr x_2\; 
 \Re\brkt{\bar{u}_{(n)}(x_2)u_{(m)}(x_2)}\cos\frac{lx_2}{R}, \nonumber\\
 g_{R(nm,l-)} \defa \frac{g}{\sqrt{\pi R}}\int\dr x_2\; 
 \Re\brkt{\bar{u}_{(n)}(x_2)u_{(m)}(x_2)}\sin\frac{lx_2}{R}, \nonumber\\
 g_{I(nm,l+)} \defa \frac{g}{\sqrt{\pi R}}\int\dr x_2\; 
 \Im\brkt{\bar{u}_{(n)}(x_2)u_{(m)}(x_2)}\cos\frac{lx_2}{R}, \nonumber\\
 g_{I(nm,l-)} \defa \frac{g}{\sqrt{\pi R}}\int\dr x_2\; 
 \Im\brkt{\bar{u}_{(n)}(x_2)u_{(m)}(x_2)}\sin\frac{lx_2}{R}, \nonumber\\
 (g^2)_{(nm,l+,p+)} \defa \frac{g^2}{\pi R}\int\dr x_2\; 
 \Re\brkt{\bar{u}_{(n)}(x_2)u_{(m)}(x_2)}\cos\frac{lx_2}{R}
 \cdot\cos\frac{px_2}{R},  \nonumber\\
 (g^2)_{(nm,l+,p-)} \defa \frac{g^2}{\pi R}\int\dr x_2\; 
 \Re\brkt{\bar{u}_{(n)}(x_2)u_{(m)}(x_2)}\cos\frac{lx_2}{R}
 \cdot\sin\frac{px_2}{R},  \nonumber\\
 (g^2)_{(nm,l-,p+)} \defa \frac{g^2}{\pi R}\int\dr x_2\; 
 \Re\brkt{\bar{u}_{(n)}(x_2)u_{(m)}(x_2)}\sin\frac{lx_2}{R}
 \cdot\cos\frac{px_2}{R},  \nonumber\\
 (g^2)_{(nm,l-,p-)} \defa \frac{g^2}{\pi R}\int\dr x_2\; 
 \Re\brkt{\bar{u}_{(n)}(x_2)u_{(m)}(x_2)}\sin\frac{lx_2}{R}
 \cdot\sin\frac{px_2}{R}. 
\eea

The first line of Eq.(\ref{minimal_g_int}) contains 
the minimal coupling of each Kaluza-Klein mode 
of the matter field~$\vph_{(n)}$ and the gauge field~$v_{(0)}^m$ 
included in $\rho_{(0)}$.

\section{Summary and discussion} \label{summary}
We derived 3D effective theory on the BPS domain wall 
which is described in terms of the 3D superfields. 
Such a superfield description of LET on the BPS wall is useful 
not only because it makes the unbroken SUSY manifest, 
but also it greatly simplifies the procedure for integrating out 
the massive modes. 
So our procedure presented in the present paper should be used 
when we derive LET on the BPS wall including interaction terms. 

The main obstacle in the derivation of LET is 
the execution of only the $\tht_1$-integration 
while leaving $\tht_2$ unintegrated. 
Thus, our main results are Eqs.(\ref{tht1intS}), 
(\ref{tht1intS2}) and (\ref{tht1intS3}). 
To obtain the 3D superfield description of the theory, 
we have to carry out the mode expansion of $\vph$, $\rho^\alpha$ and 
$\sgm$ defined by Eqs.(\ref{def_vph}) and (\ref{def_krs}). 
The mode expansion is trivial for the vector superfield 
and is given by Eq.(\ref{md_expd_V}). 
We derived the mode equation~Eq.(\ref{mode_eq}) 
in the generalized Wess-Zumino model including the case that 
the classical configuration~$\Acl(x_2)$ is complex. 
In a simple case where $\Acl(x_2)$ is real, 
it is reduced to the familiar form~Eq.(\ref{comb_md_eq})
by some field rotations. 
If we can solve the mode equation, the desired 3D LET is obtained 
by substituting the mode-expanded expression
\be
 \vph^i(x^m,x_2,\tht_2)=\Acl^i(x_2)+\frac{1}{\sqrt{2}}\sum_{n=0}^\infty 
 u_{(n)}^i(x_2)\vph_{(n)}(x^m,\tht_2)
\ee 
and Eq.(\ref{md_expd_V}) into the $\tht_1$-integrated 
actions~Eqs.(\ref{tht1intS}), (\ref{tht1intS2}) and (\ref{tht1intS3}), 
and performing the $x_2$-integration. 

There is another approach to discuss LET on the wall. 
Since the BPS wall realizes the partial SUSY breaking of 
3D $\cN=2$ to $\cN=1$, 
we can construct LET on the wall by the nonlinear realization 
approach~\cite{Ivanov}. 
This approach is useful because it does not involve the dimensional 
reduction and we do not have to suffer from integrating out 
the massive modes, since the discussion starts in three dimensions 
and only light modes are introduced from the beginning 
in this approach. 
However, since this approach uses only information about symmetries, 
we cannot obtain any information about the magnitude of various parameters 
in LET which reflect the wall structure. 
In order to discuss such parameters, 
we have to derive LET from the 4D bulk theory. 
In this sense, our derivation presented in this paper 
is a complementary approach to the nonlinear realization\footnote{
The authors of Ref.\cite{Achucarro} discussed LET on a structureless 
supersymmetric membrane. 
Their result corresponds to the thin-wall limit of ours. 
}. 

At the end of Section~\ref{real_config}, we have mentioned that 
the zero-mode~$\vph_{(0)}$ does not have self-interaction. 
By noting that $\vph_{(0)}$ in that case corresponds to 
the Nambu-Goldstone (NG) modes for the broken space-time symmetries, 
this fact can be interpreted as 
a consequence of the low-energy theorem. 
However, in the case that $\Acl(x_2)$ is complex, 
we can show that $\vph_{(0)}$ can have self-interaction. 
This seems inconsistent with the low-energy theorem. 
The cause of this contradiction is in the definition of 
the NG modes~$\vph_{(0)}$. 
In order to define the NG modes that obey the low-energy theorem, 
they must be introduced as {\it collective coordinates}. 
The introduction of NG modes for the broken space-time symmetries 
in such a way leads to the clarification of the relation between 
LET on the BPS wall in our approach and the one obtained 
by the nonlinear realization 
or the one provided in Ref.\cite{Achucarro}. 
The research along this direction is now in progress. 

For other directions of research, we would like to discuss the case that 
a domain wall is saturated the BPS bound only {\it approximately}. 
Such a situation must be considered when we try to construct 
a realistic model in the brane-world scenario, 
since our world has no exact SUSY. 
For example of such a wall configuration, 
the author has found with other collaborators 
a topologically stable non-BPS system 
that consists of an {\it approximate} BPS and anti-BPS 
domain walls in Ref.\cite{MSSS}. 
We would like to investigate whether there is a useful choice of 
the 3D superspace for the approximately preserved SUSY 
in the original 4D superspace in such a case. 
Expanding the discussion to the supergravity is also an interesting subject.


\vspace{5mm}

\begin{center}
{\bf Acknowledgments}
\end{center}
The author thanks Nobuhito Maru, Norisuke Sakai and Ryo Sugisaka 
for a collaboration of previous works which motivate this work. 

\pagebreak

\appendix
\section{Notations} \label{notations}
Basically, we follow the notations of Ref.~\cite{WB} 
for the four-dimensional bulk theory. 

\subsection{Notations for 3D theories}
The notations for the 3D theories are as follows. 

We take the space-time metric as 
\be
 \eta^{mn}=\diag(-1,+1,+1). 
\ee

The 3D $\gm$-matrices, $(\gm_{(3)}^m)_\alpha^{\;\;\beta}$, 
can be written by the Pauli matrices as  
\be
 \gm_{(3)}^0=\sgm^2, \;\;\;
 \gm_{(3)}^1=-i\sgm^3, \;\;\;
 \gm_{(3)}^3=i\sgm^1, 
\ee
and these satisfy the 3D Clifford algebra, 
\be
 \{\gm_{(3)}^m,\gm_{(3)}^n\}=-2\eta^{mn}.
\ee

The generators of the Lorentz group~$Spin(1,2)$ are 
\be
 \gm_{(3)}^{mn}\equiv\frac{1}{4}[\gm_{(3)}^m,\gm_{(3)}^n]. 
\ee

The relations between the 4D $\sgm$-matrices and 
the above $\gm_{(3)}^m$ are 
\bea
 (\sgm^\mu)_{\alpha\dot{\beta}} \eql 
 (\gm_{(3)}^0,\gm_{(3)}^1,-1,\gm_{(3)}^3)_\alpha^{\;\;\gm}
 (-\sgm^2)_{\gm\beta}, \nonumber\\
 (\bar{\sgm}^{\mu})^{\dot{\alpha}\beta} \eql 
 (-\sgm^2)^{\alpha\gm}
 (\gm_{(3)}^0,\gm_{(3)}^1,1,\gm_{(3)}^3)_\gm^{\;\;\beta}, 
 \label{sgm_gm3}
\eea
\bea
 (\sgm^{mn})_\alpha^{\;\;\beta} \eql 
 (\gm_{(3)}^{mn})_\alpha^{\;\;\beta}, \nonumber\\
 (\sgm^{m2})_\alpha^{\;\;\beta} \eql 
 \frac{1}{2}(\gm_{(3)}^m)_\alpha^{\;\;\beta}. 
\eea
Note that, in three dimensions, there is no discrimination 
between the dotted and undotted indices. 

The spinor indices are raised and lowered by multiplying $\sgm^2$ 
from the left. 
\be
 \psi_\alpha=(\sgm^2)_{\alpha\beta}\psi^\beta, \;\;\;
 \psi^\alpha=(\sgm^2)^{\alpha\beta}\psi_\beta. 
\ee
We take the following convention of the contraction of spinor indices. 
\be
 \psi_1\psi_2\equiv \psi_1^\alpha \psi_{2\,\alpha}
 =(\sgm^2)_{\alpha\beta}\psi_1^\alpha \psi_2^\beta
 =\psi_2\psi_1. 
\ee 

The relations between 4D Grassmannian coordinates~$\tht$, $\btht$ 
and 3D ones~$\tht_1$, $\tht_2$ are 
\be
 \tht^\alpha=\frac{e^{i\dlt/2}}{\sqrt{2}}(\tht_1^\alpha+i\tht_2^\alpha), 
 \;\;\;\;\;
 \btht^{\dot{\alpha}}=\frac{e^{-i\dlt/2}}{\sqrt{2}}
 (\tht_1^\alpha-i\tht_2^\alpha), 
\ee
where $\dlt$ is defined in Eq.(\ref{choice_dlt}). 
Then, it follows that 
\bea
 \tht^2 \eql e^{i\dlt}\brc{\tht_1\tht_2-\frac{i}{2}(\tht_1^2-\tht_2^2)}, 
 \nonumber\\
 \btht^2 \eql e^{-i\dlt}\brc{\tht_1\tht_2+\frac{i}{2}(\tht_1^2-\tht_2^2)}, 
 \nonumber\\ 
 \tht^2\btht^2 \eql -\tht_1^2\tht_2^2. \label{tht-thti_rel}
\eea

The definitions of the integral measures are as follows. 
\be
 \int\dr^2\tht\;\tht^2=\int\dr^2\btht\;\btht^2=1, 
\ee
and 
\be
 \int\dr^2\tht_1\;\tht_1^2=\int\dr^2\tht_2\;\tht_2^2=1. 
\ee
Then, using the relation~Eq.(\ref{tht-thti_rel}), 
the following relation is obtained. 
\be
 {\rm d}^2\tht {\rm d}^2\btht=-{\rm d}^2\tht_1 {\rm d}^2\tht_2. 
\ee

\subsection{Useful Formulae}
\bea
 (\psi_1\psi_2)^* \eql \psi_1\psi_2, \nonumber\\ 
 (\psi_1\gm_{(3)}^m \psi_2)^* \eql -\psi_1\gm_{(3)}^m \psi_2, \nonumber\\
 (\psi_1\gm_{(3)}^m \gm_{(3)}^n \psi_2)^* \eql 
 \psi_1\gm_{(3)}^m \gm_{(3)}^n \psi_2. 
\eea

\be
 \sgm^2 \gm_{(3)}^m \sgm^2=-{}^t\gm_{(3)}^m. 
\ee

\be
 \tht_1^\alpha\tht_1^\beta=-\frac{1}{2}\tht_1^2(\sgm^2)^{\alpha\beta}, \;\;\;
 \tht_{1\alpha}\tht_{1\beta}=\frac{1}{2}\tht_1^2(\sgm^2)_{\alpha\beta}.
\ee

\be
 (\tht_1\lmd)(\chi\psi)=-\frac{1}{2}
 \{(\tht_1\psi)(\chi\lmd)-(\tht_1\gm_{(3)}^m\psi)(\chi\gm_{(3)m}\lmd)\} \;\;\;
 (\mbox{Fierz transformation})
\ee

\bea
 \tht_1 \gm_{(3)}^m \tht_2 \eql -\tht_2 \gm_{(3)}^m \tht_1, \\
 \tht_1 \gm_{(3)}^m\gm_{(3)}^n \tht_2 \eql 
 \tht_2 \gm_{(3)}^n\gm_{(3)}^m \tht_1.  
\eea
In particular, 
\bea
 \tht_1 \gm_{(3)}^m \tht_1 \eql 0, \\
 \tht_1 \gm_{(3)}^m\gm_{(3)}^n \tht_1 \eql -\tht_1^2\eta^{mn}.
\eea

\section{SUSY algebra and the covariant derivatives} \label{cov_derv}
The 3D $\cN=1$ SUSY algebra is 
\be
 \{Q^{(3)}_\alpha, Q^{(3)}_\beta\}=2(\gm_{(3)}^m \sgm^2)_{\alpha\beta}P_m,
\ee
where $Q^{(3)}_\alpha$ and $P_m$ denote the supercharge and 
the translational generators. 

The representation of the generators on the 3D $\cN=1$ superspace 
$(x^m,\tht)$ is 
\bea
 \hat{P}_m \eql -i\der_m, \nonumber\\
 \hat{Q}^{(3)}_\alpha \eql \frac{\der}{\der\tht^\alpha}
 +i(\gm_{(3)}^m\tht)_\alpha \der_m.  \label{hatQ3}
\eea

For the group element $\Omega=e^{ix^m P_m+\tht Q^{(3)}}$, 
the Cartan one-form is 
\bea
 \Omega^{-1}{\rm d}\Omega \eql i({\rm d}x^m+i{\rm d}\tht\gm_{(3)}^m\tht)P_m
 +{\rm d}\tht^\alpha Q^{(3)}_\alpha \nonumber\\
 \defa i\omega_P^m P_m+\omega_Q^\alpha Q^{(3)}_\alpha. 
 \label{cartan_form1}
\eea
For the superspace coordinate differentials 
${\rm d}X^M=({\rm d}x^m,{\rm d}\tht^\alpha)$, 
the supervielbein matrix $E_M^{\;\;N}$ is defined by 
\be
 \omega^N={\rm d}X^M E_M^{\;\;N}. 
\ee
Then the covariant derivatives $\cD_N$ can be obtained by 
\be
 \cD_N=(E^{-1})_N^{\;\;M}\der_M. 
\ee
Namely, 
\bea
 \cD_m \eql \der_m, \\
 \cD_\alpha \eql \frac{\der}{\der\tht^\alpha}
 -i(\gm_{(3)}^m\tht)_\alpha \der_m. \label{3D_cvd}
\eea

By denoting the superspace generators as $\Gm_M$ collectively, 
the Cartan one-form can be expressed as follows. 
(See Eq.(\ref{cartan_form1}).)
\be
 \Omega^{-1}{\rm d}\Omega=i\omega^M \Gm_M.
\ee
Here $\Omega=e^{iX^M \Gm_M}$. 
Then 
\be
 {\rm d}\Omega=i\omega^M\Omega\Gm_M. \label{dOmega}
\ee
Noticing ${\rm d}={\rm d}X^M\der_M=\omega^M\cD_M$, Eq.(\ref{dOmega}) 
becomes 
\be
 \omega^M\cD_M \Omega=i\omega^M\Omega\Gm_M,
\ee
that is, 
\be
 \cD_M \Omega=\Omega(i\Gm_M) \label{DW_WG}
\ee
Since $\Gm_M=(P_m, -iQ^{(3)}_\alpha)$ in our case, it follows that 
\be
 \cD_\alpha \Omega=\Omega Q^{(3)}_\alpha. \label{DW_WG_ex}
\ee

\section{Orthogonality of the mode functions} \label{orthogonal}
In this appendix, we will prove the orthogonality of 
the mode functions~Eq.(\ref{orthonorm}). 
Here we assume that the mass eigenvalues of 
the mode equation~Eq.(\ref{mode_eq}) are real. 

We can easily show that 
\be
 \int\dr x_2 (\cD_y\bar{u}_{(n)i}) u_{(l)}^i
 =-\int\dr x_2\bar{u}_{(n)i} \cD_y u_{(l)}^i. 
\ee
Thus, 
\bea
 m_{(n)}\int\dr x_2 \bar{u}_{(n)i} u_{(l)}^i 
 \eql \int\dr x_2\;i\brc{\cD_y \bar{u}_{(n)i}-e^{-i\dlt}\cD_i W_j u_{(n)}^j}
 u_{(l)}^i \nonumber\\
 \eql \int\dr x_2\brc{-i\bar{u}_{(n)}^{\bar{i}}\cD_y u_{(l)\bar{i}}
 -ie^{-i\dlt}\cD_i W_j u_{(n)}^j u_{(l)}^i}. 
\eea

Using the conjugate of Eq.(\ref{mode_eq}), 
\be
 -i\brc{\cD_y u_{(l)\bar{i}}-e^{i\dlt}\bar{\cD}_{\bar{i}}
 \bar{W}_{\bar{j}}\bar{u}_{(l)}^{\bar{j}}}
 =m_{(l)}u_{(l)\bar{i}},
\ee
we can obtain 
\be
 (m_{(n)}-m_{(l)})\int\dr x_2 \bar{u}_{(n)i} u_{(l)}^i
 =-2i\Re\brc{\int\dr x_2 e^{-i\dlt}\cD_i W_j u_{(n)}^j u_{(n)}^i}. 
\ee
Therefore, in the case that $m_{(n)}\neq m_{(l)}$, we can show 
the orthogonality by taking the real part of the above equation. 
\be
 \Re\brc{\int\dr x_2 \bar{u}_{(n)i} u_{(l)}^i}=0. 
\ee

In the case that the eigenvalues $m_{(n)}$ and $m_{(l)}$ are degenerate, 
we can redefine the corresponding mode functions so that they are 
orthogonal to each other. 
In fact, if we redefine the mode function~$u_{(l)}^i(x_2)$ as 
\be
 \tl{u}_{(l)}^i(x_2)\equiv \frac{u_{(l)}^i(x_2)-C_{nl}u_{(n)}^i(x_2)}
 {1-|C_{nl}|^2}, 
\ee
where 
\be
 C_{nl}\equiv\int\dr x_2 \bar{u}_{(n)i} u_{(l)}^i, 
\ee
the functions $u_{(n)}^i(x_2)$ and $\tl{u}_{(l)}^i(x_2)$ are certainly 
orthogonal. 

As a result, by taking into account the normalization of 
each mode functions, we can obtain the desired relation, 
\be
 \Re\brc{\int\dr x_2 \bar{u}_{(n)i} u_{(l)}^i}=\dlt_{nl}.
\ee

In the above proof, we have used the assumption that all 
of the mass eigenvalues~$m_{(n)}$ are real. 
In the case that the K\"{a}hler potential is minimal, 
we can show the reality of $m_{(n)}$. 

The mode equation~Eq.(\ref{mode_eq}) can be rewritten as 
\be
 i\brc{\der_2 \bar{u}-\Gm^k_{ij}(\Acl)\der_2\Acl^j\bar{u}_k
 -e^{-i\dlt}\brkt{W_{ij}(\Acl)-\Gm^k_{ij}(\Acl)W_k(\Acl)}
 K^{j\bar{l}}(\Acl)u_{\bar{l}}}
 =m\bar{u}_i, \label{mode_eq2}
\ee
In the case of the minimal K\"{a}hler potential, 
\be
 \Gm^k_{ij}=0, \;\;\;
 K^{j\bar{l}}=\dlt^{j\bar{l}}. 
\ee
So, denoting 
\be
 \bar{\bdm{u}}\equiv \brkt{\begin{array}{c}\bar{u}_1 \\ \bar{u}_2 \\ 
 \vdots \end{array}}=\bdm{u}_{\rm R}-i\bdm{u}_{\rm I}, \;\;\;
 (\alpha_{\rm R}+i\alpha_{\rm I})_i^{\;\;\bar{l}}
 \equiv e^{-i\dlt}W_{ij}(\Acl)K^{j\bar{l}}(\Acl), 
\ee
Eq.(\ref{mode_eq2}) can be simplified as 
\be
 i\brc{\der_2\bar{\bdm{u}}-(\alpha_{\rm R}+i\alpha_{\rm I})\bdm{u}}
 =m\bar{\bdm{u}}. 
\ee
This can also be written as 
\be
 \mtrx{\alpha_{\rm I}}{\der_2+\alpha_{\rm R}}{-\der_2+\alpha_{\rm R}}
 {-\alpha_{\rm I}}\vct{\bdm{u}_{\rm R}}{\bdm{u}_{\rm I}}
 =m\vct{\bdm{u}_{\rm R}}{\bdm{u}_{\rm I}}. 
\ee
Since $\alpha_{\rm R}$ and $\alpha_{\rm I}$ are real symmetric matrices, 
the differential operator 
\be
 \cO=\mtrx{\alpha_{\rm I}}{\der_2+\alpha_{\rm R}}{-\der_2+\alpha_{\rm R}}
 {-\alpha_{\rm I}}
\ee
is hermitian. 
Therefore the eigenvalues of $\cO$, $m_{(n)}$, are real.

\end{document}